\documentclass[10pt]{article}
\usepackage{a4wide}
\usepackage{color}
\usepackage{cite}
\usepackage{graphicx}

\usepackage{amssymb}
\usepackage{amsmath}

\newcommand{\dd}{{\rm{d}}} 
\newcommand{\rovno}{\!\!\!\!& = &\!\!\!\!}
\newcommand{\F}{F} 

\def \bF {\mathbf{F}}
\def \bA {\mathbf{A}}

\begin{document}

\title{
All solutions of Einstein--Maxwell equations with \\
a cosmological constant in 2+1 dimensions
}

\author{
Ji\v{r}\'{\i} Podolsk\'y
and
Mat\'u\v{s} Papaj\v{c}\'{\i}k
\thanks{{\tt
podolsky@mbox.troja.mff.cuni.cz,
MatusPapajcik@centrum.sk
}}
\\ \ \\ \ \\
Institute of Theoretical Physics, Charles University, \\
V~Hole\v{s}ovi\v{c}k\'ach 2, 18000 Prague 8, Czech Republic.
}

\maketitle

\begin{abstract}
We present a \emph{general} solution of the coupled Einstein--Maxwell field equations (without the source charges and currents) in three spacetime dimensions. We also admit any value of the cosmological constant.  The whole family of such $\Lambda$-electrovacuum local solutions splits into two distinct subclasses, namely the non-expanding Kundt class and the expanding Robinson--Trautman class. While the Kundt class only admits electromagnetic fields which are aligned along the geometrically privileged null congruence, the Robinson--Trautman class admits both aligned and also more complex non-aligned Maxwell fields. We derive all the metric and Maxwell field components, together with explicit constraints imposed by the field equations. We also identify the most important special spacetimes of this type, namely the coupled gravitational-electromagnetic waves and charged black holes.
\end{abstract}

\vfil\noindent
PACS class:  04.20.Jb, 04.50.--h, 04.40.Nr


\vfil
\eject

\tableofcontents

\newpage

\section{Introduction}

Recently, in paper \cite{PodolskySvarcMaeda:2019} we derived the most general solution of the Einstein equations with a cosmological constant~$\Lambda$ and also an aligned pure radiation  matter field (possibly gyrating null dust/particles) in three spacetime dimensions. Here we extend this study to another important non-vacuum case, which is the presence of an electromagnetic field. In fact, we explicitly derive \emph{all} solutions of the Einstein--Maxwell field equations with any value of~$\Lambda$.

For many decades, the 2+1 dimensional Einstein gravity has attracted a lot of attention. The main reason is that such gravity theory is mathematically simpler than standard general relativity because the number of independent components of the curvature tensor is much lower. In fact, the Weyl tensor identically vanishes, and the Riemann and Ricci tensors have the same number of components. Consequently, there is \emph{no classic dynamical degree of freedom in 2+1 spacetimes}. The Ricci tensor --- directly given by the Einstein field equations --- fully determines the \emph{local curvature} of the spacetime. This implies that a general \emph{vacuum} solution of Einstein's equations is just the maximally symmetric  Minkowski, de~Sitter (dS), or anti-de~Sitter (AdS) spacetime for~${\Lambda=0}$, ${\Lambda>0}$, or~${\Lambda<0}$, respectively.

Despite such local simplicity/triviality of the 2+1 gravity theory, it can serve as a very useful playground for various investigations, ranging from the black hole properties and cosmology to high energy physics and quantum gravity. While the Einstein equations determine the spacetime locally, there can be \emph{global topological degrees of freedom} reflected in the appropriate domains of the coordinates employed: It is possible to construct globally different geometries from locally identical spacetimes by various identifications. In the context of black holes, this has been successfully used for construction of famous BTZ-type solutions with horizons when ${\Lambda<0}$ by performing nontrivial identifications of the local AdS vacuum spacetime, pure radiation solutions, or spacetimes with electromagnetic field~\cite{btz1,btz2,chargedbtz}. The corresponding topological degrees of freedom play a crucial role in quantum gravity models~\cite{carlip}. However, it is still not clear if they represent all possible non-vacuum spacetimes. It is thus desirable to obtain and investigate more general exact solutions in the presence of matter.

Many exact spacetimes in 2+1 dimensional Einstein gravity have already been found. They are nicely summarized, classified and described in a helpful comprehensive catalogue \cite{GarciaDiaz:2017}. Such solutions were found in a great number of works by making various specific assumptions on their symmetry, algebraic structure, or other geometrical or physical constraints. A general study of solutions of 2+1 dimensional Einstein--Maxwell theory using the Rainich geometrization of the electromagnetic field was presented in \cite{KrongosTorre:2017}. Using a different approach, in this paper we solve  the Einstein--Maxwell field equations generically, \emph{without making any assumption}. In fact, we systematically derive all possible such spacetimes, extending and generalizing previously known exact electrovacuum solutions.

Specifically, in Sec.~\ref{allgeometries} we recall the key result of \cite{PodolskySvarcMaeda:2019} that  (virtually) all 2+1 geometries belong either to the family of (non-expanding) Kundt spacetimes or to the family of (expanding) Robinson--Trautman  spacetimes. We also present the canonical metric form and the natural null triad. The related Appendix~\ref{sec_geomcurv} contains the corresponding Christoffel symbols and all components of the Riemann and Ricci tensors. In Sec.~\ref{genericelmagfield} we present the most general electromagnetic 2-form field in 2+1 gravity, together with its dual 1-form, the equivalent Newman--Penrose scalars, and the energy-momentum tensor. In Sec.~\ref{EinsteinMaxwell} we formulate the (source-free) Einstein--Maxwell field equations with~$\Lambda$, expressed in a simple form. Sec.~\ref{sec:Kundt} contains an explicit step-by-step integration of these field equations in the Kundt case, while Sec.~\ref{sec:RTaligned} contains an analogous procedure for the complementary Robinson--Trautman case. In both cases, the electromagnetic field is aligned with the privileged null direction of the gravitational field. The resulting complete families of such spacetimes are summarized in Subsec.~\ref{summary-Kundt} and~\ref{summary-RT}, respectively. The distinct family of Robinson--Trautman geometries with non-aligned electromagnetic fields is presented in Sec.~\ref{sec:RTnonaligned}, with a specific particular solution obtained in Subsec.~\ref{particular-solution-RT}. Final summary and further remarks can be found in concluding Sec.~\ref{sec:finalsummary}.

\newpage


\section{All geometries and their canonical form in 2+1 gravity}
\label{allgeometries}

In Sec.~2 of our previous work~\cite{PodolskySvarcMaeda:2019}, we investigated general 3-dimensional Lorentzian spacetimes $({\cal M},g_{ab})$ with the metric signature ${(++-)}$. We proved the uniqueness theorem, namely that the only possible such spacetimes are either \emph{expanding geometries} of the \emph{Robinson--Trautman type} (with ${\Theta\not=0}$) or \emph{non-expanding geometries} of the \emph{Kundt type} (with ${\Theta=0}$). They are necessarily twist-free and shear-free, see Theorem~1 in~\cite{PodolskySvarcMaeda:2019} (this observation was already made in~\cite{ChowPopeSezgin:2009}).

In a $C^1$-spacetime there exists a \emph{geodesic null} vector field~$\mathbf{k}$ (defined as a tangent vector of null geodesics at any point), which in ${D=3}$ is equivalent to \emph{hypersurface-orthogonality}, see Theorem~2 in~\cite{PodolskySvarcMaeda:2019}. Recall that the \emph{expansion}~$\Theta$ is the only nontrivial optical scalar,
\begin{equation}
\Theta = \rho \equiv  k_{a;b}\,m^a m^b \,,
\label{Theta}
\end{equation}
which characterizes the properties of a null congruence generated by~$\mathbf{k}$, in a triad ${\mathbf{e}_{I}\equiv\left\{\mathbf{k},\,\mathbf{l},\,\mathbf{m}\right\}}$
of two null vectors ${\mathbf{k}, \mathbf{l}}$ and one spatial vector $\mathbf{m}$, normalized as
\begin{equation}
\mathbf{k}\cdot\mathbf{l}=-1 \,, \qquad \mathbf{m}\cdot\mathbf{m}=1  \,.
\label{triad}
\end{equation}

In~\cite{PodolskySvarcMaeda:2019}, we also introduced \emph{canonical coordinates} ${\{r,\,u,\,x\}}$  for all Robinson--Trautman and Kundt metrics, see Theorem~3,
\begin{equation}
\dd s^2 = g_{xx}(r,u,x)\, \dd x^2+2\,g_{ux}(r,u,x)\, \dd u\, \dd x -2\,\dd u\,\dd r+g_{uu}(r,u,x)\, \dd u^2 \,. \label{general nontwist}
\end{equation}
These coordinates are adapted to their unique geometry, namely~$r$ is an affine parameter along the null congruence generated by $\mathbf{k}$, the coordinate $u$ labels null hypersurfaces (such that ${k_a \propto u_{,a}}$) which naturally foliate the spacetimes, and the spatial coordinate~$x$ spans the 1-dimensional ``transverse'' subspace with constant $u$ and $r$.

It is also convenient to recall that the non-vanishing contravariant metric components $g^{ab}$  are
\begin{equation}
g^{xx}=1/g_{xx}\,, \qquad
g^{ur}=-1\,, \qquad
g^{rx}= g_{ux}/g_{xx} \,, \qquad
g^{rr}= -g_{uu}+g_{ux}^2/g_{xx} \,, \label{CovariantMetricComponents}
\end{equation}
equivalent to the inverse relations
\begin{equation}
g_{xx}=1/g^{xx}\,, \qquad
g_{ur}=-1\,, \qquad
g_{ux}= g_{xx}\,g^{rx} \,, \qquad
g_{uu}= -g^{rr}+g_{xx}\,{(g^{rx})}^2 \,. \label{CovariantMetricComp}
\end{equation}

The most natural choice of the null triad frame ${\left\{\mathbf{k},\,\mathbf{l},\,\mathbf{m}\right\}}$ satisfying~(\ref{triad}) is
\begin{equation}
\mathbf{k}=\partial_r\,, \qquad
\mathbf{l}=\frac{1}{2}\,g_{uu}\,\partial_r+\partial_u \,, \qquad \mathbf{m}=\frac{1}{\sqrt{g_{xx}}}\,\big(g_{ux}\,\partial_r+\partial_x\big) \,.
\label{triadexpl}
\end{equation}
A direct calculation for the metric (\ref{general nontwist}) reveals that ${k_{a;b}=\frac{1}{2}g_{ab,r}}$. An explicit form of the expansion scalar~(\ref{Theta}) thus becomes ${\Theta=k_{x;x}\,m^x m^x}$, implying an important relation
\begin{equation}
g_{xx,r}=2\Theta\, g_{xx} \,,
\label{shearfree condition}
\end{equation}

For our next investigation it seems convenient to introduce a new function $G(r,u,x)$, which \emph{fully} encodes the spatial metric function ${g_{xx}>0}$ via the simple relation
\begin{equation}
G \equiv \frac{1}{\sqrt{g_{xx}}} \qquad \Leftrightarrow \qquad g_{xx} = G^{\,-2}\,.
\label{G}
\end{equation}
The key relation (\ref{shearfree condition}) then takes the form
\begin{equation}
\Theta = - (\ln G)_{,r} \,.
\label{Theta-G}
\end{equation}
Now it immediately follows that for vanishing expansion, ${\Theta=0}$, the function $G$ and thus also the spatial metric ${g_{xx}(u,x)}$ must be \emph{independent of the coordinate}~$r$. It yields the \emph{Kundt class} of non-expanding, twist-free and shear-free geometries \cite{Kundt:1961,Stephani:2003,GriffithsPodolsky:2009,PodolskyZofka:2009,OrtaggioPravdaPravdova:2013}. The complementary case ${\Theta\neq 0}$ gives the expanding \emph{Robinson--Trautman class} of geometries \cite{RobTra60,RobTra62,Stephani:2003,GriffithsPodolsky:2009,PodOrt06,OrtaggioPravdaPravdova:2013,OrtPodZof08,OrtaggioPodolskyZofka:2015}, as summarized in Theorem~4 of our work~\cite{PodolskySvarcMaeda:2019}.

The Christoffel symbols and all coordinate-components of the Riemann and Ricci curvature tensors for the general metric (\ref{general nontwist}), calculated using the relation~(\ref{shearfree condition}), are listed in Appendix~\ref{sec_geomcurv}.

\newpage


\section{Generic electromagnetic field in 2+1 gravity}
\label{genericelmagfield}

The aim of this work is to systematically investigate all possible gravitational and electromagnetic fields in 2+1 dimensions, solving the coupled Einstein--Maxwell field equations.

Based on the results summarized in previous Sec.~\ref{allgeometries}, all such spacetimes can be conveniently written in the canonical coordinates ${\{r,\,u,\,x\}}$ for the general metric~(\ref{general nontwist}). Consequently, \emph{generic electromagnetic field} takes the form of an antisymmetric ${3 \times 3}$ Maxwell tensor
\begin{equation}
F_{ab}=
\begin{pmatrix}
0 & F_{ru} & F_{rx}\\
-F_{ru} & 0 & F_{ux}\\
-F_{rx} & -F_{ux} & 0
\end{pmatrix},
\label{Fab}
\end{equation}
which is equivalent to considering the 2-form ${\bF = \tfrac{1}{2}F_{ab}\, \dd x^a \wedge \dd x^b}$, that is explicitly
\begin{equation}
\bF = F_{ru}\, \dd r \wedge \dd u + F_{rx}\, \dd r \wedge \dd x  + F_{ux}\, \dd u \wedge \dd x\,.
\label{2-formF}
\end{equation}
The field has \emph{only 3 independent components}. These can be obtained from the electromagnetic potential 1-form  ${\bA = A_{a}\, \dd x^a}$ by the standard relation
\begin{equation}
\bF = \dd \bA \,.
\label{def-potentail-A}
\end{equation}

Using (\ref{CovariantMetricComponents}), the corresponding contravariant components ${F^{ab}\equiv g^{ac}g^{bd}F_{cd}}$ read
\begin{align}
F^{ru} =-\frac{F_x}{g_{xx}}\,,\qquad
F^{rx} = \frac{F_u}{g_{xx}}\,,\qquad
F^{ux} =-\frac{F_r}{g_{xx}}\,,
\label{Fab-contra}
\end{align}
where the useful functions are
\begin{align}
F_r & \equiv F_{rx}\,, \label{Fr}\\
F_x & \equiv g_{xx}F_{ru}-g_{ux}F_{rx}\,, \label{Fx}\\
F_u & \equiv g_{ux}F_{ru}-F_{ux}-g_{uu}F_{rx}\,. \label{Fu}
\end{align}

In fact, these three functions are directly related to the components of the \emph{dual Maxwell field 1-form}  ${^*\bF = \,^*\!F_{a}\, \dd x^a}$ defined using the Hodge star operator,
\begin{equation} \label{Dual}
^*F^a \equiv \tfrac{1}{2}\,\omega^{abc}F_{bc}\,,\qquad\hbox{where}\qquad
\omega^{abc} = \frac{1}{\sqrt{-g}} \,\varepsilon^{abc}\,.
\end{equation}
Here $g$ denotes the \emph{determinant of the metric} $g_{ab}$, while $\varepsilon^{abc}$ is the completely antisymmetric Levi-Civita symbol, for which we employ the convention that
${\varepsilon^{abc}=\varepsilon_{abc} \equiv +1}$ if ${abc}$ is an even permutation of ${rux}$,
it is $-1$ for odd  permutation of ${rux}$, and $0$ otherwise. For the metric (\ref{general nontwist}) we immediately get
\begin{equation}\label{det}
-g = g_{xx} \equiv G^{\,-2}\,,
\end{equation}
and in view of \eqref{Fab} we obtain
\begin{align}
\,^*\!F^{r} =  G\,F_{ux} \,,\qquad
\,^*\!F^{u} = -G\,F_{rx} \,,\qquad
\,^*\!F^{x} =  G\,F_{ru} \,.
\label{Fa-contra}
\end{align}
Using \eqref{Fr}--\eqref{Fu}, the corresponding covariant components ${\,^*\!F_{a}  = g_{ab}\,^*\!F^{b}}$ are
\begin{align}
\,^*\!F_{a} = G\,F_{a} \,,
\label{Fa}
\end{align}
so that the dual 1-form Maxwell field reads
\begin{align}
^*\bF = G\,(\, F_{r}\, \dd r + F_{u}\, \dd u + F_{x}\, \dd x \,)\,.
\label{*Fa}
\end{align}
For completeness let us also recall the inverse relation to \eqref{Dual},
\begin{equation} \label{Dual-inverse}
F_{ab}  = - \omega_{abc}\,^*\!F^c\qquad\hbox{where}\qquad
- \omega_{abc} = \sqrt{-g} \,\,\varepsilon_{abc} = G^{\,-1}\,\varepsilon_{abc}\,.
\end{equation}

Next, it is necessary to evaluate the \emph{electromagnetic invariants}
\begin{equation}
F^2 \equiv F_{ab}\,F^{ab}\,, \qquad  \,^*\!F^2 \equiv \,^*\!F_{a} \,^*\!F^{a}\,.
\label{F2invariant-def}
\end{equation}
A direct evaluation yields
\begin{equation}
F^2 = -2\,^*\!F^2 = -2G^2 \big(g_{uu}F_{rx}^2+2F_{rx}(F_{ux}-g_{ux}F_{ru})+g_{xx}F_{ru}^2 \big)\,.
\label{F2invariant}
\end{equation}
Moreover, ${F_{ab} \,^*\!F^{a} \,^*\!F^{b} = 0 }$ due to the symmetry reasons.

Similarly as for general relativity in ${D=4}$, it is convenient to define \emph{Newman--Penrose scalars of the Maxwell field} by its three independent projections onto the frame (\ref{triadexpl}),
\begin{align}
\phi_0 &\equiv F_{ab}\,k^am^b\,,\nonumber\\
\phi_1 &\equiv F_{ab}\,k^al^b\,,\label{NPdef}\\
\phi_2 &\equiv F_{ab}\,m^al^b\,.\nonumber
\end{align}
Explicit calculation reveals that
\begin{align}
\phi_0 & = G\,F_{rx}
 && = G\,F_{r}\,, \label{NP0}\\
\phi_1 & = F_{ru}
 && = G^2\big(F_{x}+g_{ux}F_{r}\big) \,, \label{NP1}\\
\phi_2 & = G\,\big( g_{ux}F_{ru}-F_{ux}-\tfrac{1}{2}g_{uu}F_{rx}\big)
 && = G\,\big(F_{u}+\tfrac{1}{2}g_{uu}F_{r}\big)\,,\label{NP2}
\end{align}
so that the invariant can be expressed as
\begin{equation}
\tfrac{1}{2}F^2 = 2\phi_0\phi_2-\phi_1^2\,.
\label{F2invariant-scalars}
\end{equation}

These scalars have \emph{distinct boost weights} $+1$, $0$, ${-1}$, respectively, and can be used for invariant \emph{algebraic classification} of the electromagnetic field \cite{OrtaggioPravdaPravdova:2013}, based on its (non-)alignment with the geometrically privileged null vector field ${\mathbf{k}=\partial_r}$ of the metric. By definition the field is \emph{aligned} if its component with the highest boost weight vanishes. From (\ref{NP0}) we immediately observe that
\begin{align}
\hbox{\emph{electromagnetic field is aligned with}}\ \mathbf{k} \quad \Leftrightarrow  \quad
\phi _0 = 0 \quad \Leftrightarrow \quad
F_{rx} = 0  \quad \Leftrightarrow \quad
F_r =0 \,.\label{alignment}
\end{align}
It can also be shown that this is equivalent to the special property of the field, namely
\begin{equation}
F_{ab}\,k^b = {\cal N}\,k_a\,.
\label{Fisaligned}
\end{equation}
Such an aligned field has just two components, namely ${\phi _1=F_{ru}}$ and ${\phi _2 = G\,( g_{ux}F_{ru}-F_{ux})}$, and ${F^2 = -2 \phi_1^2}$. When ${\phi _1 = 0 \Leftrightarrow  F_x=0}$, the field is \emph{null}. When ${\phi _2 = 0 \Leftrightarrow  F_u=0}$, it is \emph{non-null}.

In the case when the electromagnetic field is \emph{both aligned and null}, the invariant vanishes, ${F^2=0}$. This describes \emph{purely radiative field}, i.e., a propagating electromagnetic wave characterized by the only non-vanishing component $F_{ux}$.

There is a freedom in the choice of the frame normalized as (\ref{triad}), given by the local \emph{Lorentz transformations}. It consists of a boost ${\mathbf{k}'=B\,\mathbf{k}}$,  ${\mathbf{l}'=B^{-1}\,\mathbf{l}}$ which determines the distinct boost weights $+1$, $0$, ${-1}$ of (\ref{NPdef}), respectively. The second Lorentz transformation is a null rotation with fixed $\mathbf{k}$ of the form
\begin{equation}
\mathbf{k}'=\mathbf{k}\,, \qquad
\mathbf{l}'=\mathbf{l}+\sqrt{2}L\,\mathbf{m}+L^2\,\mathbf{k}\,,\qquad
\mathbf{m}'=\mathbf{m}+\sqrt{2}L\,\mathbf{k}  \,.
\label{triad-null-rotation}
\end{equation}
There is also an analogous null rotation with fixed $\mathbf{l}$ which changes $\mathbf{k}$. However, in our case the direction of $\mathbf{k}$ is geometrically privileged (being twist-free and shear-free). Only (\ref{triad-null-rotation}) thus needs to be considered. It is easy to prove that the Maxwell scalars (\ref{NPdef}) transform as
\begin{align}
\phi_0' &= \phi_0\,,\nonumber\\
\phi_1' &= \phi_1 + \sqrt{2}L\,\phi _0\,,\label{NPdefprimed}\\
\phi_2' &= \phi_2 + \sqrt{2}L\,\phi _1 + L^2\,\phi _0\,. \nonumber
\end{align}
Of course, the expression (\ref{F2invariant-scalars}) is invariant since ${2\phi_0'\phi_2'-\phi_1'^2=2\phi_0\phi_2-\phi_1^2}$.

\newpage
Finally, we need to evaluate the \emph{energy-momentum tensor} for a generic electromagnetic field which (in any dimension, including ${D=3}$) is defined as
\begin{equation} \label{Tab}
 T_{ab} \equiv \tfrac{\kappa _0}{8\pi } \big(  F_{ac}F_b{}^c-\tfrac{1}{4}g_{ab}F^2 \big),
\end{equation}
where ${\kappa _0>0}$ is a constant depending on the choice of the physical units. Interestingly, in arbitrary dimension $D\ge 3$ the Maxwell field \emph{satisfies all the standard energy conditions}, see Proposition~21 in \cite{MaedaMartinez:2020}.

A straightforward (but somewhat lengthy) calculation reveals that
\begin{align} \label{Tabexpl}
\tfrac{8\pi}{\kappa _0}\,T_{rr} &=G^2F_{rx}^2\,, \nonumber\\
\tfrac{8\pi}{\kappa _0}\,T_{rx} &=G^2F_{rx}(g_{xx}F_{ru}-g_{ux}F_{rx})\,,\nonumber\\
\tfrac{8\pi}{\kappa _0}\,T_{ru} &=\tfrac{1}{2}G^2(g_{xx}F_{ru}^2-g_{uu}F_{rx}^2)\,,\nonumber\\
\tfrac{8\pi}{\kappa _0}\,T_{xx} &=-F_{rx}( g_{ux}F_{ru} + F_{ux})
+\tfrac{1}{2}G^2(2g_{ux}^2-g_{xx}g_{uu}) F_{rx}^2+\tfrac{1}{2}g_{xx}F_{ru}^2\,,\\
\tfrac{8\pi}{\kappa _0}\,T_{ux} &=\tfrac{1}{2}G^2 \big[ g_{ux}g_{uu}F_{rx}^2-2g_{xx}g_{uu}F_{ru}F_{rx}+g_{xx}F_{ru}(g_{ux}F_{ru}-2F_{ux})\big] \,,\nonumber\\
\begin{split}\tfrac{8\pi}{\kappa _0}\,T_{uu} &=
\tfrac{1}{2}G^2\big[ 2F_{ux}^2+2g_{uu}F_{rx}F_{ux}+g_{uu}^2F_{rx}^2-4g_{ux}F_{ru}F_{ux}\big. \nonumber\\
&\qquad \qquad \big. -2g_{ux}g_{uu}F_{rx}F_{ru}+(2g_{ux}^2-g_{xx}g_{uu})F_{ru}^2\big] \,, \end{split} \nonumber
\end{align}
and the corresponding trace ${T\equiv g^{ab}\,T_{ab}}$ is
\begin{equation} \label{traceT}
\tfrac{8\pi}{\kappa _0}\,T=G^2F_{rx}( g_{ux}F_{ru}-F_{ux}) -\tfrac{1}{2}G^2( g_{xx}F_{ru}^2 + g_{uu} F_{rx}^2)\,.
\end{equation}

Now, it is a nice fact that, by combining (\ref{Tabexpl}) with (\ref{traceT}) as ${T_{ab}-T g_{ab}}$, the result for \emph{all components} can be written in a \emph{simple factorized form} as
\begin{equation} \label{Tab-T}
\tfrac{8\pi}{\kappa _0}\,(T_{ab}-T g_{ab}) = G^2 F_a F_b\,,
\end{equation}
in terms of the functions $F_a$ encoding the electromagnetic field, which we have introduced in (\ref{Fr})--(\ref{Fu}).


\section{Einstein--Maxwell field equations with $\Lambda$}
\label{EinsteinMaxwell}

Having identified all 3-dimensional Lorentzian geometries --- which can be written in the canonical form~(\ref{general nontwist}) --- and also the generic form of the electromagnetic field (\ref{Fab}) with the energy-momentum tensor of the form (\ref{Tabexpl}) implying (\ref{Tab-T}), we can now apply the field equations.

Einstein's equations are ${R_{ab}-\frac{1}{2}R\,g_{ab}+\Lambda\, g_{ab}=8\pi\, T_{ab}}$, in which we also admit a non-vanishing \emph{cosmological constant} $\Lambda$. Their trace is ${R=2(3\Lambda-8\pi\, T)}$, so that the equations can be put into the form ${R_{ab}= 2\Lambda\,g_{ab}+8\pi\big(T_{ab}-T\,g_{ab}\big)}$. For the generic electromagnetic field $F_{ab}$ we have derived the nice relation (\ref{Tab-T}), and thus the \emph{Einstein field equations} in 2+1 gravity with $\Lambda$, coupled to an electromagnetic field, are simply
\begin{equation}
R_{ab} = 2\Lambda\,g_{ab} + \kappa _0\,G^2 F_a F_b \,, \label{EinstinEq}
\end{equation}
where the functions $F_a$ are defined by (\ref{Fr})--(\ref{Fu}). Expressed in terms of the dual Maxwell field $^*\bF$ 1-form components, see \eqref{*Fa} and  \eqref{Fa}, these are even simpler, namely
\begin{equation}
R_{ab} = 2\Lambda\,g_{ab} + \kappa_0 \,^*\!F_{a}\!\,^*\!F_{b} \,. \label{EinstinEq-dual}
\end{equation}

In addition to these equations for the gravitational field represented by the metric $g_{ab}$, we must also satisfy the \emph{Maxwell equations} ${\dd^*\bF=4\pi\,^*{\mathbf{J}}}$ and ${\dd\bF=0}$ for the electromagnetic field  $F_{ab}$. In the \emph{absence of electric charges and currents}, in components these read ${F^{ab}{}_{;b}=0}$, ${F_{[ab;c]}=0}$. They are equivalent to
\begin{align}
(\sqrt{-g}\,F^{ab})_{,b} &=0\,, \label{Maxeq1}\\
 F_{[ab,c]} &=0\,, \label{Maxeq2}
\end{align}
where, using \eqref{det},
\begin{equation}
\sqrt{-g} = \sqrt{g_{xx}} = G^{\,-1}\,.
\end{equation}

Recall also that
the source-free Maxwell equation ${\dd^*\bF=0}$, which is equivalent to \eqref{Maxeq1}, in components reads ${\,^*\!F_{[a,b]}=0}$. In view of \eqref{Fa}, it can be directly written as
\begin{align}
(G\,F_{a})_{,b} = \big(G\,F_{b})_{,a}\,.
\label{Maxeq-dual}
\end{align}

Our task is to completely integrate the coupled system of the field equations (\ref{EinstinEq}) and (\ref{Maxeq1}),  (\ref{Maxeq2}) [or, equivalently, (\ref{Maxeq-dual}) instead of (\ref{Maxeq1})] in 2+1 dimensions for (\ref{general nontwist}) and (\ref{Fab}), both for the non-expanding Kundt spacetimes (Sec.~\ref{sec:Kundt}) and the expanding Robinson--Trautman spacetimes (Sec.~\ref{sec:RTaligned} and Sec.~\ref{sec:RTnonaligned}). Explicit components of the Ricci tensor $R_{ab}$, which enter (\ref{EinstinEq}), for these twist-free and shear-free geometries are given by equations (\ref{Ricci rr})--(\ref{Ricci uu}) in Appendix~\ref{sec_geomcurv}.

\subsection{Einstein field equations with a massless scalar field}
\label{Einstein-scalar}

Let us also remark that in three dimensions there is a relation between the Einstein--Maxwell system (\ref{EinstinEq-dual}) and the Einstein gravity equations (minimally) coupled to a \emph{massless scalar field}~$\Phi$ such that
\begin{equation}
g^{ab}\,\Phi_{;ab}  = 0 \,. \label{scalar}
\end{equation}
Indeed, the corresponding energy-momentum tensor reads
\begin{equation} \label{Tab-scalar}
 T_{ab} \equiv \Phi_{,a}\,\Phi_{,b} - \tfrac{1}{2} g_{ab}\,\Phi_{,c}\,\Phi^{,c}\,,
\end{equation}
implying the trace ${T = -\tfrac{1}{2}\,\Phi_{,c}\,\Phi^{,c} }$, so that the Einstein equations ${R_{ab}= 2\Lambda\,g_{ab}+8\pi\big(T_{ab}-T\,g_{ab}\big)}$ become
\begin{equation}
R_{ab} = 2\Lambda\,g_{ab} + 8\pi \,\Phi_{,a}\,\Phi_{,b} \,. \label{EinstinEq-scalar}
\end{equation}
With the identification
\begin{equation}
 \Phi_{,a} \equiv \sqrt{\frac{\kappa_0}{8\pi}}\, \,^*\!F_{a}\,, \label{identification}
\end{equation}
this system of equations is clearly equivalent to (\ref{EinstinEq-dual}).
The dual Maxwell field 1-form is thus
\begin{equation}
^*\bF = \sqrt{\frac{8\pi}{\kappa_0}}\,\dd \Phi\,.
\label{def-potential-Phi}
\end{equation}

\newpage


\section{All Kundt solutions}
\label{sec:Kundt}

In this section, we explicitly perform a step-by-step integration of the field equations in the non-expanding case ${\Theta=0}$, which defines the Kundt family of spacetimes. Recall a consequence of (\ref{G}) and (\ref{Theta-G}), namely that the function $G$ is now $r$-independent. It can be renamed as ${G(u,x)\equiv P(u,x)}$. Also the 1-dimensional spatial metric ${g_{xx}=G^{-2}}$ must be $r$-independent, that is
\begin{equation}
g_{xx}\equiv P^{-2}(u,x) \,. \label{KSpMetr}
\end{equation}
Of course, ${g^{xx}=P^2}$. Now, we will employ the Einstein field equations (\ref{EinstinEq}), which for the Kundt spacetimes take the form
\begin{equation}
R_{ab} = 2\Lambda\,g_{ab} + \kappa _0 P^2 F_aF_b \,. \label{KEinstinEq}
\end{equation}

\subsection{Integration of ${R_{rr} = \kappa _0 P^2 F_r^2}$}
In view of Eq.~(\ref{Ricci rr}), ${R_{rr} = 0}$ for ${\Theta=0}$. Therefore, this Einstein equation immediately requires ${F_r=0}$, that is
\begin{equation}\label{KFrx}
F_{rx}=0\,.
\end{equation}
It means that, inevitably, \emph{any electromagnetic field in the 2+1 Kundt spacetimes must be aligned with $\mathbf{k}=\partial _r$.} Such fields are fully described by the functions
\begin{align}
F_r = 0\,,\qquad
F_x = P^{-2}F_{ru}\,, \qquad
F_u = g_{ux}F_{ru}-F_{ux}\,. \label{KFrFxFu}
\end{align}
There are only \emph{two} possible components of the electromagnetic field, namely $F_{ru}$ and  $F_{ux}$.

In fact, this result is analogous to the situation in standard 3+1 general relativity, for which it is well known that (due to the Mariot--Robinson theorem) any Einstein--Maxwell field (including a cosmological constant $\Lambda$) in the Kundt class of geometries must be aligned, see the introductions to Chapter~31 of~\cite{Stephani:2003} and Chapter~18 of~\cite{GriffithsPodolsky:2009}.

\subsection{Integration of ${R_{rx} = \kappa _0 P^2 F_r F_x }$}
The Ricci tensor component (\ref{Ricci rp}) for ${\Theta=0}$ reduces to ${R_{rx}=-\frac{1}{2}g_{ux,rr}}$. Since ${F_r=0}$, we obtain a general solution of this Einstein equation
\begin{equation}
g_{ux}= e(u,x) + f(u,x)\,r \,, \label{KNediagCov}
\end{equation}
where $e$ and $f$ are arbitrary functions of $u$ and $x$.
In view of Eqs.~(\ref{CovariantMetricComponents}) and (\ref{KSpMetr}), the corresponding contravariant component of the Kundt metric is
\begin{equation}
g^{rx}=P^2\big[ e(u,x) + f(u,x)\,r \big] \,. \label{KNediagContra}
\end{equation}

\subsection{Integration of ${R_{ru}= -2\Lambda + \kappa _0 P^2 F_r F_u}$}
Using Eqs.~(\ref{KSpMetr}) and (\ref{KNediagCov}), the Ricci tensor component (\ref{Ricci ru}) is
${R_{ru} = -\frac{1}{2}\, g_{uu,rr} +\frac{1}{2} P^2(f_{||x}+f^2)}$, where
\begin{equation}
f_{||x} \equiv f_{,x} + \frac{P_{,x}}{P}\,f
\quad\Leftrightarrow\quad Pf_{||x} \equiv (Pf)_{,x}\,.
\label{f||x}
\end{equation}
Actually, the symbol ${\,_{||}}$ denotes the covariant derivative (of a 1-form $f$) related to the spatial metric $g_{xx}$ on the 1-dimensional ``transverse'' subspace with constant $u$ and $r$, namely  ${f_{||x} = f_{,x}-\,^{S}\Gamma^{x}_{xx}\,f}$,
where ${\,^{S}\Gamma^x_{xx}\equiv\tfrac{1}{2}g^{xx}g_{xx,x}}$ is the corresponding Christoffel symbol (see Appendix~\ref{sec_geomcurv}). Although this  notation seems to be superficial here, we employ it in order to see the  relation to our previous studies \cite{{PodSva13,PodSva15,PodSva16}} of Kundt and Robinson--Trautman spacetimes in any higher dimension ${D\ge4}$ where this geometric notation plays a key role.

Because ${F_r=0}$, the corresponding Einstein equation thus simplifies, and can be integrated to
\begin{equation}
g_{uu}= a(u,x)+b(u,x)\,r + c(u,x) \,r^2 \,, \label{KguuExpl}
\end{equation}
where $a(u,x)$ and $b(u,x)$ are \emph{arbitrary} functions, while
\begin{equation}
c(u,x) \equiv 2\Lambda+{\textstyle\frac{1}{2}} P^2(f_{||x}+f^2) \,. \label{Kdefc}
\end{equation}

\subsection{Integration of the Maxwell equations}
The crucial $r$-dependence of all metric functions for the 2+1 Kundt spacetimes is thus determined. In general, $g_{uu}$ is quadratic, $g_{ux}$ is linear, and ${g_{xx}\equiv P^{-2}(u,x)}$ is independent of $r$. Now, applying the Maxwell equations (\ref{Maxeq1}), (\ref{Maxeq2}) with ${\sqrt{-g} = P^{-1}}$, we will determine the $r$-dependence of the electromagnetic field.

In the present setting, there are only 4 independent Maxwell equations , namely 3 components of
${(\sqrt{-g}\,F^{ab})_{,b}=0}$ and just 1 component of ${F_{[ab,c]}=0}$. Because (\ref{Fab-contra}) with (\ref{KFrFxFu}) implies
\begin{align}
F^{ru} =-F_{ru}\,,\qquad
F^{rx} = P^2(g_{ux}F_{ru}-F_{ux})\,,\qquad
F^{ux} = 0\,,
\label{KFab-contra}
\end{align}
these 4 equations for the electromagnetic field have the form
\begin{align}
F_{ru,r} &= 0 \,,\label{K-Maxwel-1}\\
(g_{ux}F_{ru}-F_{ux})_{,r} &= 0 \,,\label{K-Maxwel-2}\\
\big(P(g_{ux}F_{ru}-F_{ux})\big)_{,x}  &= \Big(\frac{F_{ru}}{P}\Big)_{,u}\,,\label{K-Maxwel-3}\\
F_{ux,r}+F_{ru,x} &= 0 \,.\label{K-Maxwel-4}
\end{align}
These equations can be completely solved for the two non-trivial components $F_{ru}$ and $F_{ux}$. Starting with (\ref{K-Maxwel-1}), we immediately obtain that
\begin{equation}\label{KFru}
F_{ru} = Q(u,x)\,,
\end{equation}
where ${Q(u,x)}$ is an arbitrary function independent of $r$. By employing (\ref{K-Maxwel-4}), we thus get
\begin{equation}\label{KFux}
F_{ux} = -Q_{,x}\,r - \xi(u,x)\,,
\end{equation}
where ${\xi(u,x)}$ is another arbitrary function. Equation (\ref{K-Maxwel-2}) gives the constraint
\begin{equation}\label{KFconstraint-1}
Q_{,x} = - f\,Q\,,
\end{equation}
and (\ref{K-Maxwel-3}) reduces to the equation
\begin{equation}\label{KFconstraint-2}
\big(P(e\,Q + \xi)\big)_{,x} = \Big(\frac{Q}{P}\Big)_{,u} \,.
\end{equation}

To summarize, by integrating all the Maxwell equations we obtained explicit components of the (necessarily aligned) \emph{electromagnetic field in any 2+1 Kundt spacetime},
\begin{align}
F_{rx} = 0 \,,\qquad
F_{ru} = Q \,,\qquad
F_{ux} = f\,Q\,r - \xi \,,
\label{KFab-explicit}
\end{align}
where the functions $Q(u,x)$ and $\xi(u,x)$ are constrained by equations (\ref{KFconstraint-1}), (\ref{KFconstraint-2}). Consequently,
\begin{align}
F_r = 0\,,\qquad
F_x = P^{-2}\,Q\,, \qquad
F_u = e\,Q + \xi \,, \label{KFrFxFuexplicit}
\end{align}
and, due to~(\ref{NP0})--(\ref{NP2}),
\begin{align}
\phi _0 = 0 \,, \qquad
\phi _1 = Q \,, \qquad
\phi _2 = P \,( e\,Q + \xi ).\label{K-NP012}
\end{align}
When ${\phi _1 = 0 \Leftrightarrow  Q=0}$, the field is \emph{null}, and then ${\phi _2 = P\,\xi}$. When ${\phi _2 = 0 \Leftrightarrow  e\,Q = -\xi}$, it is \emph{non-null}, and then ${\phi _1 = Q}$.

Now, we can integrate the remaining three Einstein equations, which impose the unique relation between the gravitational and electromagnetic field components.

\subsection{Integration of ${R_{xx}=2\Lambda\,g_{xx} + \kappa _0 P^2 F_x^2}$}
For ${\Theta=0}$, using Eqs.~(\ref{fxx}) and (\ref{KNediagCov}), the Ricci tensor component (\ref{Ricci pq}) reduces to $R_{xx}= -f_{xx}\equiv-\big(f_{||x}+ \frac{1}{2}f^2\big)$.
The field equation ${R_{xx}=2\Lambda\,g_{xx}+ \kappa _0 P^2 (P^{-2}\,Q)^2 = (2\Lambda + \kappa _0 \,Q^2)\,P^{-2}}$ implies
\begin{equation}
 \kappa _0 \,Q^2 = - \big[\, 2\Lambda + P^2(f_{||x}+ \tfrac{1}{2} f^2) \,\big]\,. \label{K-QfLambdaP}
\end{equation}
The electromagnetic field component ${F_{ru} \equiv \phi _1 = Q(u,x)}$ is thus \emph{explicitly determined by the cosmological constant $\Lambda$ and by the metric functions $P, f$} (provided the right-hand side of (\ref{K-QfLambdaP}) is non-negative). It is now convenient to introduce a rescaled form of $f$ entering the metric function ${g_{ux}= e + f\,r}$, see (\ref{KNediagCov}), namely
\begin{equation}
\F \equiv P^2f^2\,.
\label{defF}
\end{equation}
Then the field equation (\ref{K-QfLambdaP}) can be rewritten as
\begin{equation}
 P^2(f_{||x}+f^2) =  {\textstyle\frac{1}{2}} F - 2\Lambda - \kappa _0\,Q^2 \,. \label{K-identity}
\end{equation}
We can thus simplify the metric function ${g_{uu}}$, namely its coefficient $c$ in (\ref{KguuExpl}) given by (\ref{Kdefc}), to
\begin{equation}
c(u,x) = \Lambda + \tfrac{1}{4} F - \tfrac{\kappa _0}{2}Q^2 \,. \label{Kdefc-alter}
\end{equation}
At this stage, the most general Kundt solution in ${D=3}$ takes the form
\begin{eqnarray}
\dd s^2 \rovno \frac{\dd x^2}{P^2} +2\,(e+f\,r)\,\dd u \dd x -2\,\dd u\dd r  +\Big[a+b\,r +\big(\Lambda+\tfrac{1}{4} F - \tfrac{ \kappa _0 }{2} Q^2 \big)\,r^2 \Big]\, \dd u^2 \,, \label{K-Kundtmetric}
\end{eqnarray}
and the Einstein--Maxwell field equation (\ref{K-identity}) using (\ref{f||x}) reads
\begin{equation}
 P (Pf)_{,x} =  - ( 2\Lambda + \tfrac{1}{2} F + \kappa _0\,Q^2) \,. \label{K-QfLambdaPgen}
\end{equation}

\subsection{Integration of ${R_{ux}=2\Lambda\,g_{ux} + \kappa _0 P^2 F_u F_x}$}
Eq.~(\ref{Ricci up}) with ${\Theta=0}$ for the metric (\ref{K-Kundtmetric}) gives $ R_{ux} = \frac{1}{2}\big[f_{,u}-b_{,x}-e P^2 (f_{||x}+f^2)-f\,(\ln P)_{,u}\big] - \frac{1}{4}\big[(F - 2\kappa _0 Q^2)_{,x}
+2 f P^2(f_{||x}+f^2)\big]\,r$. Applying (\ref{K-identity}) and (\ref{KFrFxFuexplicit}), (\ref{KNediagCov}), the corresponding field equation
${R_{ux}=2\Lambda\,g_{ux} + \kappa _0 Q (e\,Q + \xi)=2\Lambda e + \kappa _0  (e\,Q^2 + Q\xi) + 2\Lambda f\,r}$ splits into two conditions, resulting from the coefficients for the powers $r^1$ and $r^0$, namely
\begin{align}
F_{,x} - 2\kappa _0 (Q^2)_{,x}
+ (F - 4\Lambda - 2 \kappa _0\,Q^2)f   & = -8\Lambda f \,,\label{K-Rux-1}\\
f_{,u}-b_{,x}-(\tfrac{1}{2} F - 2\Lambda - \kappa _0\,Q^2 )e-f\,(\ln P)_{,u}  & = 4\Lambda e + 2\kappa _0  (e\,Q^2 + Q\xi) \,.\label{K-Rux-2}
\end{align}
Using the field equation (\ref{K-QfLambdaPgen}), Eq.~(\ref{K-Rux-1}) simplifies to
${(Q^2)_{,x} =- 2 Q^2 f}$ which is \emph{identically satisfied} due to (\ref{KFconstraint-1}). Only the constraint (\ref{K-Rux-2}) thus remains, which can be put into the form
\begin{align}
b_{,x} =  f_{,u} - f\,(\ln P)_{,u} - \tfrac{1}{2} (F+4\Lambda+2\kappa _0 Q^2)\, e - 2\kappa _0 Q\,\xi \,,\label{K-Rux-final}
\end{align}
that is
\begin{align}
b_{,x} =   P \,\Big(\frac{f}{P}\Big)_{,u} + Pe\, (Pf)_{,x} - 2\kappa _0 Q\,\xi \,.\label{K-Rux-final-alternatively}
\end{align}
This is an explicit expression \emph{determining the metric function} $b(u,x)$.

\subsection{Integration of ${R_{uu}=2\Lambda\,g_{uu} + \kappa _0 P^2 F_u^2}$}
For ${\Theta=0}$ and the Kundt metric (\ref{K-Kundtmetric}), using the relation ${e_{||x}\equiv e_{,x}+e\,P_{,x}/P}$ and similar for ${f_{||x}}$, ${e_{,u||x}}$, ${f_{,u||x}}$, ${a_{||xx}}$, ${b_{||xx}}$ and ${c_{||xx}}$ (see Appendix~\ref{sec_geomcurv}),  the last Ricci tensor component (\ref{Ricci uu}) reads
\begin{equation}
R_{uu}= A+B\,r+C\,r^2\,,
\label{RuuKundt}
\end{equation}
where
\begin{eqnarray}
A \rovno  a(c-\tfrac{1}{2}F)
+ P^2 \Big[
-\tfrac{1}{2}a_{,xx}+\tfrac{1}{2}a_{,x}\Big(f-\frac{P_{,x}}{P}\Big)
-\tfrac{1}{2}b\Big(e_{,x}+\frac{P_{,x}}{P}\,e+ \frac{P_{,u}}{P^3}\Big)  \nonumber\\
&& \hspace{24mm}
+\big(f_{,u}-b_{,x}-c\,e\big)e+\Big(e_{,ux}+ \frac{P_{,x}}{P}\,e_{,u}\Big)
+ \frac{P_{,uu}}{P^3}-2\frac{P_{,u}^2}{P^4}\Big]\,,
\label{A}\\
B \rovno b\big(c-\tfrac{1}{2}F-\tfrac{1}{2} P (Pf)_{,x}\big)
+ P^2 \Big[  \big(f_{,u}-\tfrac{1}{2}b_{,x}\big)_{,x}
+ (f_{,u}-\tfrac{1}{2} b_{,x})\Big(f+\frac{P_{,x}}{P}\Big)
\nonumber\\
&& \hspace{44mm}
-c\,\Big(e_{,x}+\frac{P_{,x}}{P}\,e+ \frac{P_{,u}}{P^3}\Big)
-2e\,(c_{,x} +f\,c)  \Big] \,,
\label{B}\\
C \rovno c(c-F)
- P^2\Big[ \,\tfrac{1}{2}c_{,xx}+\tfrac{1}{2} c_{,x}\Big(3f+\frac{P_{,x}}{P}\Big)
+c \Big(f_{,x}+\frac{P_{,x}}{P}f + \tfrac{1}{2}f^2\Big)\Big]\,.
\label{C}
\end{eqnarray}
Due to (\ref{KguuExpl}), (\ref{KFrFxFuexplicit}), the corresponding field equation is ${R_{uu} = 2\Lambda(a+b\,r + c\,r^2 ) + \kappa _0 P^2 (e\,Q + \xi)^2 }$, which splits into the following three constraints
\begin{align}
A &= 2\Lambda\,a + \kappa _0 P^2 (e\,Q + \xi)^2 \,,\label{K-Ruu-1}\\
B &= 2\Lambda\,b \,,\label{K-Ruu-2}\\
C &= 2\Lambda\,c \,.\label{K-Ruu-3}
\end{align}

From (\ref{Kdefc-alter}), (\ref{K-QfLambdaPgen}), (\ref{KFconstraint-1}) we easily derive interesting identities for spatial derivatives of $c$,
\begin{align}\label{cx-cxx}
c_{,x}=-f\,c\,, \qquad    c_{,xx}=(f^2-f_{,x})\,c\,.
\end{align}

By using (\ref{cx-cxx}), the expression (\ref{C}) reduces to
${C = c\big[c-\tfrac{1}{2}F-\tfrac{1}{2}P(Pf)_{,x}\big]}$, and substituting from (\ref{Kdefc-alter}), (\ref{K-QfLambdaPgen}) we obtain ${C = 2\Lambda\,c}$. The equation (\ref{K-Ruu-3}) \emph{is thus identically satisfied}.

Surprisingly, the equation (\ref{K-Ruu-2}) \emph{is also identically satisfied}. Applying (\ref{K-QfLambdaPgen}), the first term in (\ref{B}) yields ${2\Lambda\,b}$, while the complicated combination of various terms in the square brackets vanishes by using  the relations (\ref{cx-cxx}), (\ref{K-Rux-final}), (\ref{Kdefc-alter}) and the field equations (\ref{KFconstraint-1}), (\ref{KFconstraint-2}). Therefore, ${B=2\Lambda b}$, which is the equation (\ref{K-Ruu-2}).

We are thus left with \emph{only one equation}, namely (\ref{K-Ruu-1}). Using (\ref{K-QfLambdaP}), (\ref{Kdefc-alter}), (\ref{K-QfLambdaPgen}) and (\ref{K-Rux-final}), it can be simplified to
\begin{align}
a_{,xx}   - & a_{,x}\Big(f-\frac{P_{,x}}{P}\Big) -a\Big(f_{,x}+\frac{P_{,x}}{P}f\Big) \nonumber\\
=& -b\Big(e_{,x}+\frac{P_{,x}}{P}\,e+\frac{P_{,u}}{P^3}\Big)+2\Big(e_{,ux}+ \frac{P_{,x}}{P}\,e_{,u}\Big) \label{K-Ruu-a}\\
 & -Pe^2(Pf)_{,x}+2ef\frac{P_{,u}}{P} + 2\bigg( \frac{P_{,uu}}{P^3}-2\frac{P_{,u}^2}{P^4}\bigg) -2\kappa _0\,\xi ^2\,. \nonumber
\end{align}
This equation \emph{determines the last metric function} $a(u,x)$.

Alternatively, it can be understood as an \emph{explicit expression for the} $\xi(u,x)$ component of the Maxwell field, in terms of the metric functions ${P, e, f, a, b}$. Such an equation can be expressed in a covariant form as
\begin{align}
2\kappa _0\,\xi ^2  =& - a_{||xx} + (fa)_{||x} -b\Big(e_{||x}+\frac{P_{,u}}{P^3}\Big)+2(e_{,u})_{||x} \label{K-Ruu-a-covar}\\
 & -P^2e^2 f_{||x}+2ef\frac{P_{,u}}{P} + 2\bigg( \frac{P_{,uu}}{P^3}-2\frac{P_{,u}^2}{P^4}\bigg) \,, \nonumber
\end{align}
where ${a_{||xx} \equiv a_{,xx} + \frac{P_{,x}}{P}\,a_{,x}}$ and
${\psi_{||x} \equiv \psi_{,x}+\psi\,P_{,x}/P}$, for  $\psi$ representing $a_{,x}$, $f$, $e$ and $e_{,u}$.


\newpage

\subsection{Summary of the Kundt solutions}
\label{summary-Kundt}
We have thus solved all the Einstein--Maxwell equations with a cosmological constant~$\Lambda$ in 2+1 gravity  for the complete Kundt family of non-expanding spacetimes. The generic gravitational field of this type is
\begin{align}
g_{xx} &= P^{-2}(u,x)\,,  \nonumber\\
g_{ux} &= e(u,x)+f(u,x)\,r \\
g_{uu} & = a(u,x)+b(u,x)\,r+c(u,x)\,r^2\,, \nonumber
\end{align}
where
\begin{align}
c = \Lambda + \tfrac{1}{4} F - \tfrac{ \kappa _0 }{2} Q^2 \,, \label{Kcsummary}
\end{align}
with
\begin{align}
F \equiv P^2f^2  \,, \label{KdefF}
\end{align}
c.f. (\ref{Kdefc-alter}), (\ref{defF}),while the electromagnetic field (\ref{KFab-explicit}) reads
\begin{align}
F_{rx} &= 0\,,  \nonumber\\
F_{ru} &= Q(u,x) \\
F_{ux} &= f(u,x)\,Q(u,x)\,r-\xi (u,x)\,. \nonumber
\end{align}
Written explicitly in a compact form,
\begin{align}
\dd s^2 =&\
\frac{\dd x^2}{P^2} + 2\,(e+f\,r)\, \dd u\, \dd x -2\,\dd u\,\dd r \nonumber \\
& +\Big(a+b\,r+\big( \Lambda + \tfrac{1}{4} F - \tfrac{ \kappa _0 }{2} Q^2 \big)\,r^2\Big)\, \dd u^2 \,, \label{Kmetric-final}
\end{align}
and
\begin{equation}
\bF =  Q\, \dd r \wedge \dd u + (f\,Q\,r-\xi)\, \dd u \wedge \dd x \,,
\label{KformF-final}
\end{equation}
corresponding to the potential
\begin{equation}
\bA = A_r\,\dd r + A_x\,\dd x \,,
\label{KformArAx-final}
\end{equation}
where, considering (\ref{KFconstraint-1}),
\begin{equation}
{\textstyle A_r \equiv -\int \! Q\, \dd u \,, \qquad  A_x \equiv r \int \! f\,Q\, \dd u  - \int \! \xi\, \dd u } \,.
\label{KformA-final}
\end{equation}

It is now important to recall the \emph{Maxwell scalars} given by (\ref{K-NP012}),
\begin{align}
\phi_0 &= 0\,,  \nonumber\\
\phi_1 &= Q\,, \label{K-NP012-summary}\\
\phi_2 &= P \,( e\,Q + \xi ) \,. \nonumber
\end{align}
We have thus proved that \emph{all electromagnetic fields in the Kundt spacetimes} in 2+1 gravity \emph{are necessarily aligned} (${\phi _0=0}$). Moreover, they split into \emph{two distinct subclasses}:

\begin{itemize}

\item \textbf{The case} ${\phi _1 = 0 \Leftrightarrow  Q=0}$: The field is \emph{null}, in which case ${\phi _2 = P\,\xi}$ and ${F_{ux}=-\xi}$, so that
\begin{equation}
\bF =  - \xi\, \dd u \wedge \dd x \,.
\label{KformF-final-null}
\end{equation}

\item \textbf{The case} ${\phi _2 = 0 \Leftrightarrow  \xi = -e\,Q}$:
The field is \emph{non-null} with only ${\phi _1 = Q}$, corresponding to
\begin{equation}
\bF =  Q\, \dd r \wedge \dd u + Q\,(e+f\,r)\, \dd u \wedge \dd x \,.
\label{KformF-final-nonnull}
\end{equation}

\end{itemize}

Notice also that, applying the Lorentz null rotation (\ref{triad-null-rotation}) with fixed $\mathbf{k}$ and the uniquely chosen parameter ${L=-\frac{1}{\sqrt{2}}\,e\,P}$ in (\ref{NPdefprimed}), the scalars (\ref{K-NP012-summary}) transform to
\begin{align}
\phi_0' &= 0\,,\nonumber\\
\phi_1' &= Q\,,\label{KNPdefprimed}\\
\phi_2' &= P\,\xi\,. \nonumber
\end{align}
Therefore, with respect to the triad with
${\mathbf{m}'=\mathbf{m}+\sqrt{2}L\,\mathbf{k} = P\,(\partial_x  + f\,r\, \partial_r)}$, the condition for the Maxwell field being non-null is ${\phi _2' = 0 \Leftrightarrow  \xi = 0}$.

The \emph{two electromagnetic components} ${Q, \xi}$ and the \emph{five metric functions} ${P, e, f, a, b}$ describing the gravitational field are mutually constrained by the following Einstein--Maxwell field equations:
\begin{align}
Q_{,x} &= - f\,Q\,, \label{KFconstraint-summary1}\\
(QPe + P\xi)_{,x} &= \Big(\frac{Q}{P}\Big)_{,u}\,, \label{KFconstraint-summary2}\\
P (Pf)_{,x} &=  - ( 2\Lambda + \tfrac{1}{2} F + \kappa _0\,Q^2)  \label{KFconstraint-summary3}\,,\\
b_{,x} &=   P \,\Big(\frac{f}{P}\Big)_{,u} + Pe\, (Pf)_{,x}-2\kappa _0 Q\,\xi\,,
\label{KFconstraint-summary4}\\
a_{,xx}   -  a_{,x}&\Big(f-\frac{P_{,x}}{P}\Big) -a\Big(f_{,x}+\frac{P_{,x}}{P}f\Big) \nonumber\\
&= -b\Big(e_{,x}+\frac{P_{,x}}{P}\,e+\frac{P_{,u}}{P^3}\Big)+2\Big(e_{,ux}+ \frac{P_{,x}}{P}\,e_{,u}\Big) \label{KFconstraint-summary5}\\
 & \quad -Pe^2(Pf)_{,x}+2ef\frac{P_{,u}}{P} + 2\bigg( \frac{P_{,uu}}{P^3}-2\frac{P_{,u}^2}{P^4}\bigg) -2\kappa _0\,\xi ^2\,. \nonumber
\end{align}
see Eqs.~(\ref{KFconstraint-1}), (\ref{KFconstraint-2}), (\ref{K-QfLambdaPgen}), (\ref{K-Rux-final-alternatively}) and (\ref{K-Ruu-a}).

Interestingly, the form of the electromagnetic field (\ref{KformF-final}) and also the same field equations (\ref{KFconstraint-summary1})--(\ref{KFconstraint-summary5}) can \emph{formally} be obtained by setting ${D=3}$ in the corresponding equations for \emph{higher-dimensional} Kundt spacetimes with an aligned Maxwell field \cite{PodolskyZofka:2009}.

\vspace{2mm}
Let us now separately discuss \emph{two geometrically distinct subclasses}, namely ${f=0}$ and ${f\ne 0}$.

\subsubsection{The subclass ${f=0}$}

From (\ref{KdefF}) it follows that ${f=0 \Leftrightarrow F=0}$, so that the equations (\ref{KFconstraint-summary1})--(\ref{KFconstraint-summary5}) considerably simplify to
\begin{align}
Q_{,x} &= 0\,, \label{KFconstraint-summary1-f=0}\\
(QPe + P\xi)_{,x} &= \Big(\frac{Q}{P}\Big)_{,u}\,, \label{KFconstraint-summary2f=0}\\
\kappa _0\,Q^2 &= -2\Lambda\,,   \label{KFconstraint-summary3f=0}\\
b_{,x} &= -2\kappa _0 Q\,\xi\,,  \label{KFconstraint-summary4f=0}\\
(P a_{,x} )_{,x}
&= -b\Big((Pe)_{,x}+\frac{P_{,u}}{P^2}\Big)  + 2(Pe_{,u})_{,x}
+ 2\Big(\frac{P_{,u}}{P^2}\Big)_{,u}   -2\kappa _0\,P\,\xi ^2\,. \label{KFconstraint-summary5f=0}
\end{align}
In this case, $Q$ \emph{is necessarily a constant}, and ${\Lambda\le0}$ because
\begin{equation}
2\Lambda = - \kappa _0 \,Q^2\,. \label{K-Q-f=0}
\end{equation}
Therefore, the electromagnetic component ${\phi_1}$ is also independent of~$u$ and~$x$,
\begin{align}
F_{ru} = \phi_1 = Q = \sqrt{-\tfrac{2}{\kappa_0}\,\Lambda}\,. \label{KQ-for-f=0}
\end{align}

Keeping both the functions $P(u,x)$ and $\xi(u,x)$ \emph{arbitrary}, the equation \eqref{KFconstraint-summary2f=0} determines the metric function $e(u,x)$. Moreover, the function $b(u,x)$ is directly determined by the spatial integral of  $\xi $ via (\ref{KFconstraint-summary4f=0}). Finally, integrating \eqref{KFconstraint-summary5f=0} we obtain $a(u,x)$.

Thus, we have obtained a complete and explicit family of such electrovacuum Kundt spacetimes in 2+1 gravity, namely
\begin{align}
\dd s^2 =&\
\frac{\dd x^2}{P^2} + 2\,e\, \dd u\, \dd x -2\,\dd u\,\dd r +\big(a+b\,r+2\Lambda\,r^2\big)\, \dd u^2 \,, \label{Kmetric-final-f=0}
\end{align}
and
\begin{equation}
\bF =  Q\, \dd r \wedge \dd u -\xi\, \dd u \wedge \dd x \,.
\label{KformF-final-f=0}
\end{equation}

It admits \emph{four physically distinct subcases}:

\begin{itemize}

\item \textbf{The case} ${Q = 0 = \xi}$: The electromagnetic field $\bF$ \emph{vanishes}, and necessarily ${\Lambda=0}$. The metric is
\begin{align}
\dd s^2 =&\
\frac{\dd x^2}{P^2} + 2\,e\, \dd u\, \dd x -2\,\dd u\,\dd r +\big(a+b\,r\big)\, \dd u^2 \,, \label{Kmetric-final-f=0-case-A}
\end{align}
where $b(u)$ is independent of $x$. It is a vacuum solution without a cosmological constant, and thus in 2+1 gravity it must be \emph{flat Minkowski space}. We derived this metric in our previous work \cite{PodolskySvarcMaeda:2019}, see Eq.~(82) with ${\mathcal{J}=0=\mathcal{N}}$ therein.

\item \textbf{The case} ${Q = 0}$: Again, ${\Lambda=0}$ and ${b=b(u)}$, so that the metric has the form (\ref{Kmetric-final-f=0-case-A}), but there is now a \emph{radiative (null) electromagnetic field}
\begin{equation}
\bF =  -\xi\, \dd u \wedge \dd x \,.
\label{KformF-final-null2}
\end{equation}
The amplitude $\xi(u,x)$ must satisfy the field equation (\ref{KFconstraint-summary2f=0}), which is ${(P\,\xi)_{,x} = 0}$. Therefore,
\begin{equation}
\xi(u,x) = \frac{\gamma(u)}{P(u,x)}\,,
\label{KformF-final-null-xi}
\end{equation}
where $\gamma(u)$ is \emph{an arbitrary profile function} of the retarded time~$u$.
Finally, $a(u,x)$ is then obtained by integrating the remaining field equation (\ref{KFconstraint-summary5f=0}).

\item \textbf{The case} ${\xi = 0}$:
The \emph{electromagnetic field is non-null}, and has the form
\begin{equation}
\bF =  Q\, \dd r \wedge \dd u \,,
\label{KformF-final-nonnull2}
\end{equation}

where $Q$ is a \emph{constant uniquely determined by negative cosmological constant} ${\Lambda}$ via (\ref{KQ-for-f=0}). The electromagnetic field is thus uniform, and positive (or zero) $\Lambda$ is not allowed.

The metric is of the form (\ref{Kmetric-final-f=0}). The field equation (\ref{KFconstraint-summary4f=0}) implies that ${b=b(u)}$, while the remaining (\ref{KFconstraint-summary2f=0}) and (\ref{KFconstraint-summary5f=0}) reduce to
\begin{align}
(Pe)_{,x}  &= -\frac{P_{,u}}{P^2} \,, \label{KFconstraint-summary2f=0,xi=0}\\
(P a_{,x})_{,x} &= 2(Pe_{,u})_{,x} - 2(Pe)_{,ux} \,. \label{KFconstraint-summary5f=0,xi=0}
\end{align}
The latter can be immediately integrated to
\begin{align}
a_{,x} =  2e_{,u} - \frac{2}{P}\big(Pe\big)_{,u} + \frac{\delta(u)}{P} \,, \label{KFconstraint-summary5f=0,xi=0integ}
\end{align}
where $\delta(u)$ is any function of $u$. After prescribing an arbitrary metric function $P(u,x)$, we obtain $e(u,x)$ by integrating (\ref{KFconstraint-summary2f=0,xi=0}), and $a(u,x)$ by integrating (\ref{KFconstraint-summary5f=0,xi=0integ}).

\item \textbf{The general case} ${Q \ne 0}$, ${\xi \ne 0}$: In the generic case with both the non-null component of the electromagnetic field  ${Q=\hbox{const.}}$ and its null component $\xi(u,x)$, we obtain the superposition (\ref{KformF-final-f=0}). The metric reads (\ref{Kmetric-final-f=0}), with a cosmological constant ${\Lambda<0}$ (notice that ${\Lambda=0}$ implies ${Q=0}$ due to (\ref{K-Q-f=0}), while ${\Lambda>0}$ is forbidden). The metric functions $a$ and~$b$ are determined by the differential equations (\ref{KFconstraint-summary4f=0}) and (\ref{KFconstraint-summary5f=0}), respectively, and there is also the constraint (\ref{KFconstraint-summary2f=0}) determining $e$.

    This family of  Kundt spacetimes in 2+1 gravity can be interpreted as \emph{mutually coupled exact gravitational and electromagnetic waves} (characterized by the functions $a(u,x)$ and $\xi(u,x)$, respectively) which propagate on the \emph{background with ${\Lambda<0}$ and uniform Maxwell field} (characterized by the constant $Q$).
    The simplest such background is
    \begin{align}
    \dd s^2 = \dd x^2 -2\,\dd u\,\dd r + 2\Lambda\,r^2\, \dd u^2 \,, \label{Kmetric-final-f=0-background}
    \end{align}
    which is the 2+1 analogue of the exceptional electrovacuum type D metric with ${\Lambda<0}$
    found by Pleba\'nski and Hacyan \cite{PlebanHacyan:1979}, see also Eq.~(7.20) in \cite{GriffithsPodolsky:2009}. Indeed, introducing ${{\cal U}=1/(2\Lambda u)}$ and ${{\cal V}=2(u+1/(\Lambda r))}$, the metric (\ref{Kmetric-final-f=0-background}) takes the form ${\dd s^2 = \dd x^2 -2\,\dd {\cal U}\dd {\cal V}/(1-\Lambda\, {\cal U}{\cal V})^2}$ which is clearly the direct-product $E^1\times$AdS$_2$ spacetime.

\end{itemize}

\subsubsection{The subclass ${f\ne 0}$}

Recalling ${F \equiv P^2f^2}$, cf. (\ref{KdefF}), in this case ${F\not=0}$. The Kundt metric takes the general form (\ref{Kmetric-final}), the aligned electromagnetic field is (\ref{KformF-final}), and the corresponding Einstein--Maxwell field equations are (\ref{KFconstraint-summary1})--(\ref{KFconstraint-summary5}).

By inspecting this system, it is seen that the first three differential equations (\ref{KFconstraint-summary1}), (\ref{KFconstraint-summary2}), (\ref{KFconstraint-summary3}) relate the metric functions $P,e,f$ and the electromagnetic field components $Q,\xi$. Subsequently, the remaining two equations (\ref{KFconstraint-summary4}) and (\ref{KFconstraint-summary5}) can be used to evaluate the metric functions $b$ and $a$, respectively.

Starting with (\ref{KFconstraint-summary1}), we immediately observe that there are \emph{two distinct subcases}:

\begin{itemize}

\item \textbf{The case} ${Q = 0}$: The electromagnetic field is \emph{null} (with ${\phi _1 = 0, \phi _2 = P\,\xi}$),
\begin{equation}
\bF =  - \xi\, \dd u \wedge \dd x \,.
\label{KformF-final-null-fnot=0}
\end{equation}
The field equation (\ref{KFconstraint-summary1}) is identically satisfied, putting no restriction on the function~$f$, while (\ref{KFconstraint-summary2}), (\ref{KFconstraint-summary3}) reduce to
\begin{align}
P\,\xi &= \gamma(u)\,, \label{KFconstraint-summary2-Q=0}\\
P (Pf)_{,x} &=  - \big( 2\Lambda + \tfrac{1}{2} (Pf)^2 \big)  \label{KFconstraint-summary3Q=0}\,.
\end{align}
The first equation determines $\xi$, giving the  same expression as (\ref{KformF-final-null-xi}), i.e., ${\xi(u,x)=P^{-1}\,\gamma(u)}$, while the second equation can be integrated for the variable $(Pf)$ in terms of the integral of $P^{-1}$, yielding
\begin{equation}
f(u,x) = -2\sqrt{\Lambda}\,P^{-1}\tan [\,\sqrt{\Lambda}\,{\textstyle\int} P^{-1}\,\dd x]\qquad\hbox{for}\ \Lambda>0\,,
\label{KFconstraint-summary3Q=0-int}
\end{equation}
and the expression for ${\Lambda<0}$ is analogous, replacing $\tan$ by $\tanh$.

In the final step, the metric functions $b$ and $a$ are obtained by integrating the field equations (\ref{KFconstraint-summary4}) and (\ref{KFconstraint-summary5}), respectively.

\item \textbf{The case} ${Q \ne 0}$:
In this generic case, the field equation (\ref{KFconstraint-summary1}) explicitly determines the metric function $f$ in terms of the electromagnetic field component $Q$, which occurs in
\begin{equation}
\bF =  Q\, \dd r \wedge \dd u + (f\,Q\,r-\xi)\, \dd u \wedge \dd x \,,
\label{KformF-again}
\end{equation}
as
\begin{equation}
f(u,x)=-(\ln Q)_{,x}\,. \label{K-QfLambdaPgen-fne0-f}
\end{equation}
However, there is a further constraint given the field equation (\ref{KFconstraint-summary3}),
\begin{align}
P (Pf)_{,x} &=  - \big( 2\Lambda + \tfrac{1}{2} (Pf)^2 + \kappa _0\,Q^2 \big)\,.
  \label{KFconstraint-summary3Qne0}
\end{align}

Notice that it can also be rewritten as
\begin{equation}
F_{,x} = - f\,(F+4\Lambda + 2 \kappa _0 \,Q^2)\,, \label{K-QfLambdaPgen-fne0}
\end{equation}
or, equivalently,
\begin{equation}
 \kappa _0 \,Q^2 = - \frac{1}{2f} \Big[\, \F_{,x} + (F+4\Lambda)f  \,\Big]\,. \label{K-Q-fnot=0}
\end{equation}
It remains to be investigated what are the constraints resulting from the simultaneous solution of equations  (\ref{K-QfLambdaPgen-fne0-f}) and (\ref{K-Q-fnot=0}).

\end{itemize}

\newpage


\section{All aligned Robinson--Trautman solutions}
\label{sec:RTaligned}

After completing the derivation and preliminary description of the non-expanding Kundt class, we will now concentrate on systematic integration of the field equations in the \emph{expanding} case ${\Theta\not=0}$, which defines the Robinson--Trautman family of spacetimes.

Recall that the field equations (\ref{EinstinEq}) take the form
\begin{equation}
R_{ab} = 2\Lambda\,g_{ab} + \kappa _0\,G^2 F_a F_b \,, \label{RTEinstinEq}
\end{equation}
where $F_a$ are defined by (\ref{Fr})--(\ref{Fu}). In this section \emph{we assume that the electromagnetic field is aligned} with ${\mathbf{k}=\partial_r} $, see (\ref{alignment}), that is
\begin{align}
F_{rx} = 0  \quad \Leftrightarrow \quad
F_r =0 \,.\label{RTalignment}
\end{align}
This considerably simplifies the field equations (\ref{RTEinstinEq}) whenever at least one of the index ${a,b}$ is $r$.

\subsection{Integration of ${R_{rr} = 0}$}
From Eq.~(\ref{Ricci rr}) we immediately get the constraint
\begin{equation}
\Theta_{,r}+\Theta^2=0 \,, \label{feqrr}
\end{equation}
which determines the $r$-dependence of the expansion scalar $\Theta$. Its general solution can be written as $\Theta^{-1}=r+r_{0}(u,x)$. Because the metric (\ref{general nontwist}) is invariant under the gauge transformation ${r \to r-r_{0}(u,x)}$, without loss of generality we can set the integration function ${r_{0}(u,x)}$ to zero. The expansion thus simplifies to
\begin{equation}
\Theta=\frac{1}{r}\,. \label{ExplEx}
\end{equation}
Integrating now the key relation (\ref{Theta-G}) we obtain
\begin{equation}
G(r,u,x) = \frac{P(u,x)}{r} \,,
\label{GRTaligned}
\end{equation}
where $P(u,x)$ is any function independent of~$r$. Using (\ref{G}), we immediately get the generic spatial metric function ${g_{xx}\equiv G^{-2}}$ in the form
\begin{equation}
g_{xx}=\frac{r^2}{P^2(u,x)} \,. \label{SpMetric}
\end{equation}
Of course, by inversion  ${g^{xx}=P^2\,r^{-2}}$.

\subsection{Integration of ${R_{rx}= 0}$}
Using Eqs.~(\ref{Ricci rp}) and (\ref{feqrr}), which implies Eq.~(\ref{ExplEx}), the Ricci tensor component ${R_{rx}}$ becomes
\begin{equation}
{\textstyle R_{rx}=-\frac{1}{2}\left(g_{ux,rr} - g_{ux,r}\,r^{-1} \right) } \,.
\label{feqrp}
\end{equation}
The corresponding field equation ${R_{rx}= 0}$ can  be integrated, yielding a general solution
\begin{equation}
g_{ux}= e(u,x)\,r^2+f(u,x) \,, \label{NediagCov}
\end{equation}
where $e$ and $f$ are arbitrary functions of $u$ and $x$.
In view of Eqs.~(\ref{CovariantMetricComp}) and (\ref{SpMetric}), the contravariant component of the Robinson--Trautman metric is
\begin{equation}
g^{rx}=P^2\big[e(u,x)+f(u,x)\,r^{-2} \big] \,. \label{NediagContra}
\end{equation}

\subsection{Integration of the Maxwell equations}
Now, applying the Maxwell equations (\ref{Maxeq1}), (\ref{Maxeq2}) with ${\sqrt{-g} = G^{-1} =r/P}$, we will determine the electromagnetic field. There are only 4 independent Maxwell equations, namely 3 components of
${(\sqrt{-g}\,F^{ab})_{,b}=0}$ and just 1 component of ${F_{[ab,c]}=0}$. Because (\ref{Fab-contra}) with (\ref{RTalignment}) implies
\begin{align}
F^{ru} =-F_{ru}\,,\qquad
F^{rx} = \frac{P^2}{r^2}\,(g_{ux}F_{ru}-F_{ux})\,,\qquad
F^{ux} = 0\,,
\label{RTFab-contra}
\end{align}
these 4 equations for the electromagnetic field take the form
\begin{align}
(r\,F_{ru})_{,r} &= 0 \,,\label{RT-Maxwel-1}\\
\big(\, r^{-1}(g_{ux}F_{ru}-F_{ux})\big)_{,r} &= 0 \,,\label{RT-Maxwel-2}\\
r^2\Big(\frac{F_{ru}}{P}\Big)_{,u} &= \Big(P(g_{ux}F_{ru}-F_{ux})\Big)_{,x} \,,\label{RT-Maxwel-3}\\
F_{ux,r}+F_{ru,x} &= 0 \,.\label{RT-Maxwel-4}
\end{align}
They can be solved for the non-trivial components $F_{ru}$ and $F_{ux}$. From (\ref{RT-Maxwel-1}) we get
\begin{equation}\label{RTFru}
F_{ru} = \frac{Q(u,x)}{r}\,,
\end{equation}
where ${Q(u,x)}$ is an arbitrary function of $u$ and $x$. By employing (\ref{RT-Maxwel-4}), we thus obtain
\begin{equation}\label{RTFux}
F_{ux} = -Q_{,x}\,\ln|r| - \xi(u,x)\,,
\end{equation}
where ${\xi(u,x)}$ is another arbitrary function. Equation (\ref{RT-Maxwel-2}) with (\ref{NediagCov}) then reduces to
\begin{equation}\label{RTFconstraint-1}
\Big(\frac{f\,Q}{r^2} + Q_{,x}\,\frac{\ln|r|}{r} + \frac{\xi}{r} \Big)_{,r}=0\,,
\end{equation}
which gives the following three independent constraints
\begin{equation}\label{RTFconstraint-1abs}
f\,Q=0\,,\qquad
Q_{,x}=0\,,\qquad
\xi=Q_{,x}\,,
\end{equation}
so that ${\xi=0}$ and ${Q=Q(u)}$ is independent of $x$.

We thus conclude that the components of a generic \emph{aligned electromagnetic field in any 2+1 Robinson--Trautman spacetime} can be written as
\begin{align}
F_{rx} = 0 \,,\qquad
F_{ru} = \frac{Q(u)}{r} \,,\qquad
F_{ux} = 0 \,,
\label{RTFab-explicit}
\end{align}
with the constraint
\begin{equation}\label{RTFconstraint-fQ=0}
f\,Q=0\,,
\end{equation}
and the Maxwell equation (\ref{RT-Maxwel-3}) which reduces to
\begin{equation}\label{RTFconstraint-2}
\Big(\frac{Q}{P}\Big)_{,u} = Q\,(e\,P)_{,x} \,.
\end{equation}

Consequently,
\begin{align}
F_r = 0\,,\qquad
F_x = P^{-2}\,Q\,r\,, \qquad
F_u = e\,Q\,r  \,, \label{RTFrFxFuexplicit}
\end{align}
and, due to~(\ref{NP0})--(\ref{NP2}),
\begin{align}
\phi _0 = 0 \,, \qquad
\phi _1 = \frac{Q}{r} \,, \qquad
\phi _2 = e\,P\,Q\,.\label{RT-NP012}
\end{align}
When ${\phi _1 = 0 \Leftrightarrow  Q=0}$ then ${\phi _2 = 0}$. Therefore, there are \emph{no null electromagnetic fields of this type}. When ${\phi _2 = 0 \Leftrightarrow  e\,Q = 0}$, it is \emph{non-null}, and then ${\phi _1 = Q(u)/r}$. Notice also, that due to (\ref{RTFconstraint-fQ=0}), either we have a vacuum solution (${Q=0}$) or a non-null electromagnetic field characterized by $Q(u)$ in the Robinson-Trautman spacetime \emph{without the non-diagonal metric term} (${g_{ux}=0}$).

\newpage
Now, we will integrate the remaining Einstein's equations which couple the gravitational and electromagnetic fields. In view of (\ref{RTFconstraint-fQ=0}), there are two cases to consider, namely ${Q=0}$ and ${f=0}$.

\begin{itemize}

\item \textbf{The case} ${Q=0}$: The electromagnetic field completely vanishes, so that the spacetimes are \emph{vacuum} (with any cosmological constant $\Lambda$). All such Robinson--Trautman solutions in 2+1 gravity were found and described in our previous work \cite{PodolskySvarcMaeda:2019}. Interestingly, for these vacuum spacetimes the function $f$ remains non-vanishing (which is not true in ${D\ge4}$).

\item \textbf{The case} ${f=0}$: In this case, the metric component $g_{ux}$ reduces to
\begin{equation}
g_{ux}= e\,r^2 \qquad \Leftrightarrow \qquad
g^{rx}=P^2\, e\,. \label{Nediagf=0}
\end{equation}
This simplifies the generic Ricci tensor components in Appendix~\ref{sec_geomcurv}, which will now apply.

\end{itemize}

\subsection{Integration of ${R_{ru}= -2\Lambda}$}
Using~(\ref{Nediagf=0}), (\ref{ExplEx}) and (\ref{SpMetric}), the Ricci tensor component (\ref{Ricci ru}) becomes
\begin{eqnarray}
&&  R_{ru} =
 -\tfrac{1}{2}\big( r\, g_{uu,r}\big)_{,r}\,r^{-1}
 +\tfrac{1}{2} c\,r^{-1} +2\,P^2e^2 \,, \label{Rru EEq}
\end{eqnarray}
where
\begin{equation}
c \equiv 2P^2\big(e_{||x}-\tfrac{1}{2} h_{xx,u}\big)\,,\qquad
e_{||x} \equiv e_{,x} + e\,P_{,x}/P \,, \label{defeparder}
\end{equation}
from which we obtain useful identities
\begin{equation}
P\,e_{||x} = (Pe)_{,x} \,, \qquad
e\,P^2\,e_{||x} = {\textstyle \frac{1}{2}}(P^2e^2)_{,x} \,,  \label{idno2}
\end{equation}
and thus
\begin{equation}
 c = 2\big[P(Pe)_{,x}+(\ln P)_{,u}\big]\,.  \label{idno3}
\end{equation}
With Eq.~(\ref{Rru EEq}), the  Einstein equation ${R_{ru}= -2\Lambda}$ can now be easily integrated to give
\begin{equation}
g_{uu}=-a-b\,\ln |r|+c\,r
+(\Lambda+P^2e^2)\,r^2 \,, \label{guuExpl}
\end{equation}
where $a(u,x)$ and $b(u,x)$ are arbitrary functions.
The $r$-dependence of all metric components is thus fully established.

\subsection{Integration of ${R_{xx}=2\Lambda\,g_{xx} + \kappa_0\,G^2 F_x^2}$}
Using Eqs.~(\ref{feqrr})--(\ref{SpMetric}) and (\ref{Nediagf=0}), the general Ricci tensor component (\ref{Ricci pq}) becomes
\begin{equation}
 R_{xx}= -c\,P^{-2}\,r -2\,e^2\,r^{2} + P^{-2}\,r\,g_{uu,r} \,.
\end{equation}
Substituting now the expression (\ref{guuExpl}), we obtain
${R_{xx} = 2\Lambda\,g_{xx}-b/P^2}$. The corresponding Einstein equation with (\ref{RTFrFxFuexplicit}) reads ${R_{xx}=2\Lambda\,g_{xx}+ \kappa_0 Q^2/P^2}$. It is satisfied if, and only if,
\begin{equation}
b(u)=-\kappa _0\, Q^2 \,. \label{bje0}
\end{equation}

\subsection{Integration of ${R_{ux}=2\Lambda\,g_{ux} + \kappa_0\,G^2 F_u F_x}$}
Using Eqs.~(\ref{ExplEx}), (\ref{SpMetric}), (\ref{Nediagf=0}), and (\ref{guuExpl}) with (\ref{bje0}), the Ricci tensor component ${R_{ux}}$ given by Eq.~(\ref{Ricci up}) reads
\begin{equation}
R_{ux}= 2\Lambda\,g_{ux} + \kappa_0\, e\, Q^2 -\tfrac{1}{2}\,a_{,x}\,r^{-1}\,.
\end{equation}
The field equation with (\ref{RTFrFxFuexplicit}) is ${R_{ux}=2\Lambda\,g_{ux} + \kappa_0\, e\, Q^2}$, so that we obtain just one simple constraint
\begin{equation}
 a_{,x} = 0\quad \Leftrightarrow \quad a=a(u) \,.
 \label{fieleq_ux}
\end{equation}
The function $a$ can depend only on the coordinate~$u$, and the most general Robinson--Trautman aligned electrovacuum solution thus takes the form
\begin{align}
\dd s^2 =&\  \frac{r^2}{P^2}\, \dd x^2+2\,e\,r^2\,\dd u \dd x -2\,\dd u\dd r \nonumber \\
& +\Big(-a(u) +\kappa_0\,Q^2(u)\ln |r|  +2\big[ P(Pe)_{,x}+(\ln P)_{,u} \big]\,r +(\Lambda+P^2e^2)\,r^2\Big)\, \dd u^2 \,. \label{RTmetric-prelim}
\end{align}

\subsection{Integration of ${R_{uu}=2\Lambda\,g_{uu} + \kappa_0\,G^2 F_u^2}$}
The Ricci tensor component ${R_{uu}}$ for the metric (\ref{RTmetric-prelim}), given generally by Eq.~(\ref{Ricci uu}), becomes
\begin{equation}
R_{uu}=  2\Lambda\,g_{uu} + A
+ \tfrac{1}{2}\big[\,a_{,u}-(a-\tfrac{1}{2}b)c - \triangle c \,\big]\frac{1}{r}
+ \tfrac{1}{2}\big[\,b_{,u}-b c\,\big]\frac{\ln r}{r} \,,
\label{Ruu1}
\end{equation}
where
\begin{equation}
 A = -P^2e^2\, b+\tfrac{1}{4}c^2 + \tfrac{1}{2}P^2e\,c_{,x} - \tfrac{1}{2} c_{,u}
   - \tfrac{1}{2}\triangle(P^2e^2) + P(Pe_{,u})_{,x} - 2\frac{P_{,u}^2}{P^2} + \frac{P_{,uu}}{P} \,,
 \label{RTdefA}
\end{equation}
$c$ is given by Eq.~(\ref{idno3}), and
\begin{equation}
\triangle c \equiv h^{xx}\,c_{||xx} = P(Pc_{,x})_{,x}
\label{Laplace}
\end{equation}
is the covariant Laplace operator on the 1-dimensional transverse Riemannian space spanned by~$x$, applied on the function~$c$. Remarkably, after substitution from (\ref{idno3}) and evaluation, the expression for $A$ enormously simplifies to
\begin{equation}
 A = -P^2e^2\, b \,.
 \label{RT-Asimplified}
\end{equation}
Moreover, using the Maxwell equation (\ref{RTFconstraint-2}) which can be rewritten as
\begin{equation}\label{RTFconstraint-2c}
Q_{,u} = \tfrac{1}{2}c\,Q\,,
\end{equation}
and the relation (\ref{bje0}), that is ${b=-\kappa _0\, Q^2}$, we easily prove that ${b_{,u}=b\,c}$. The last term in (\ref{Ruu1}) thus always vanishes. To summarize, the last Ricci tensor component takes the form
\begin{equation}
R_{uu}=  2\Lambda\,g_{uu} + \kappa _0\, e^2P^2Q^2
+ \tfrac{1}{2}\big[\,a_{,u}-(a-\tfrac{1}{2}b)c - \triangle c \,\big]\frac{1}{r} \,.
\label{Ruu1-simplified}
\end{equation}
Using (\ref{RTFrFxFuexplicit}), the corresponding field equation reads ${R_{uu}=2\Lambda\,g_{uu} + \kappa_0 \,e^2 P^2 Q^2}$, so that we obtain \emph{only one additional condition} determined by the term proportional to ${r^{-1}}$, namely
\begin{eqnarray}
 a_{,u}\rovno (a+\tfrac{\kappa_0}{2}\,Q^2)\,c + \triangle c \,. \label{RTEq2}
\end{eqnarray}

Let us observe that the equation (\ref{RTFconstraint-2c}) implies
\begin{equation}\label{RTFconstraint-consequence}
 c(u) = 2\, (\ln Q)_{,u}\,,
\end{equation}
i.e., the function $c$ must necessarily be independent of the spatial coordinate~$x$. Due to (\ref{Laplace}), ${\triangle c =0}$, and the field equation (\ref{RTEq2}) reduces to
\begin{equation}
 a_{,u} = (a+\tfrac{\kappa_0}{2}\,Q^2)\,c \,. \label{RTEq2+}
\end{equation}
Its general solution with (\ref{RTFconstraint-consequence}) is
\begin{equation}
 a(u) = Q^2\big(\kappa_0\ln |Q| - \mu\big)\,, \label{a(u)-explicit}
\end{equation}
where $\mu$ is any constant. The metric function $a(u)$ is thus directly related to the electromagnetic field $Q(u)$.

\subsection{Summary of the aligned Robinson--Trautman solutions}
\label{summary-RT}
We have solved all the Einstein--Maxwell equations with a cosmological constant~$\Lambda$ and \emph{aligned} electromagnetic field in 2+1 gravity for the Robinson--Trautman family of expanding spacetimes. In the canonical coordinates, the generic gravitational field of this type is
\begin{align}
g_{xx} &=P^{-2}(u,x)\,r^2\,,  \nonumber\\
g_{ux} &=e(u,x)\,r^2\,, \\
g_{ur} &=-1\,,  \nonumber\\
g_{uu} &= \mu\, Q^2(u) - \kappa_0\,Q^2\ln \Big|\frac{Q}{r}\Big|
  + 2\,(\ln Q)_{,u}\,r +(\Lambda+P^2e^2)\,r^2\,, \nonumber
\end{align}
where $\mu$ is a constant, $Q(u)$ is any function of $u$, and the metric functions ${P, e}$ satisfy the field equation (\ref{RTFconstraint-2}), that is
\begin{equation}\label{RTFconstraint-2Q}
\Big(\frac{Q}{P}\Big)_{,u} = Q\,(e\,P)_{,x} \,.
\end{equation}
The corresponding aligned electromagnetic field reads
\begin{align}
F_{rx} &= 0\,,  \nonumber\\
F_{ru} &= \frac{Q(u)}{r}\,, \\
F_{ux} &= 0\,, \nonumber
\end{align}
see (\ref{RTFab-explicit}), i.e., it  has \emph{only one component} $F_{ru}$.

Written explicitly in the usual compact form, the solution is
\begin{align}
\dd s^2 =&\  \frac{r^2}{P^2}\,
\big( \dd x + e \,P^2 \dd u \big)^2 -2\,\dd u\dd r \nonumber \\
& +\Big(\mu\, Q^2 - \kappa_0\,Q^2 \ln \Big|\frac{Q}{r}\Big|
  +2\,(\ln Q)_{,u}\,r + \Lambda\,r^2\Big)\, \dd u^2  \label{RTmetric-final}
\end{align}
with
\begin{equation}
\bF =  \frac{Q}{r}\, \dd r \wedge \dd u
\qquad \hbox{equivalent to} \qquad
^*\bF = \frac{Q}{P}\, \dd x + e\,P\,Q\, \dd u \,,
\label{RTformF-final}
\end{equation}
corresponding to the potential
\begin{equation}
 \bA =  Q \,\ln\frac{r}{r_0}\,  \dd u \,,
\label{RTformA-final}
\end{equation}
and the Maxwell scalars (\ref{RT-NP012})
\begin{align}
\phi_0 &= 0\,,  \nonumber\\
\phi_1 &= \frac{Q}{r}\,, \\
\phi_2 &= e\,P\,Q\,. \nonumber
\end{align}
It follows that there are \emph{no aligned (purely) null  electromagnetic fields} in the Robinson--Trautman spacetimes in 2+1 gravity because ${\phi _1 = 0}$ implies ${\phi _2 = 0}$. Moreover, ${\phi _2 = 0 \Leftrightarrow  e\,Q = 0}$. Either we have a vacuum solution (${Q=0}$) or a non-null electromagnetic field characterized by $Q(u)$ in the Robinson-Trautman spacetime without the non-diagonal metric term $g_{ux}$ (${e=0}$).

\textbf{The simplest}  ${e \ne 0}$ solution of the field equation (\ref{RTFconstraint-2Q}), which can be rewritten as
\begin{equation}\label{RT-field-equation}
(\ln P)_{,u} + P\,(e\,P)_{,x} = (\ln Q)_{,u} \,,
\end{equation}
is
\begin{equation}\label{RT-simplest}
P=1\,,\qquad
e= x\,(\ln Q)_{,u} + \alpha(u) \,,
\end{equation}
where $\alpha(u)$ is an arbitrary function of~$u$, yielding the metric
\begin{align}
\dd s^2 =&\  r^2\,
\Big( \dd x + \big( \alpha + x\,(\ln Q)_{,u}\big) \,\dd u \Big)^2 -2\,\dd u\dd r \nonumber \\
& +\Big(\mu\, Q^2 - \kappa_0\,Q^2 \ln \Big|\frac{Q}{r}\Big|
  +2\,(\ln Q)_{,u}\,r + \Lambda\,r^2\Big)\, \dd u^2\,.  \label{RTmetric-simplest}
\end{align}

\newpage
Another interesting subclass of the Robinson--Trautman spacetimes (\ref{RTmetric-final}) with aligned Maxwell field (\ref{RTformF-final}) arises when both sides of the field equation (\ref{RTFconstraint-2Q}) vanish,
${(Q/P)_{,u}=0 \Leftrightarrow (e\,P)_{,x}=0}$. Then the metric functions $P$ and $e$ are \textbf{both factorized} in the coordinates $u$ and $x$ as
\begin{equation}
 P=Q(u)\,\beta(x)\,,\qquad
 e=\frac{\alpha(u)}{Q(u)\,\beta(x)}\,, \label{Pe-explicit}
\end{equation}
where $\alpha(u)$, $\beta(x)$ are arbitrary functions of the respective coordinates. Consequently, ${e\,P=\alpha(u)}$.  (For  ${\beta=1}$ we obtain simply ${P(u)=Q(u)}$.) In such a case, the metric (\ref{RTmetric-final}) takes the form
\begin{align}
\dd s^2 =&\  \frac{r^2}{Q^2}\,
\Big( \frac{\dd x}{\beta}+\alpha\,Q\, \dd u\Big)^2 -2\,\dd u\dd r \nonumber \\
& +\Big(\mu\, Q^2 - \kappa_0\,Q^2 \ln \Big|\frac{Q}{r}\Big|
  +2\,(\ln Q)_{,u}\,r + \Lambda\,r^2\Big)\, \dd u^2\,,  \label{RTmetric-final-factorized}
\end{align}
and the Maxwell scalars are
\begin{align}
\phi_0 = 0\,,  \qquad
\phi_1 = \frac{Q}{r}\,, \qquad
\phi_2 = \alpha \,Q \,. \nonumber
\end{align}
With respect to the natural triad (\ref{triadexpl}), there are thus two components of the admitted Maxwell field, namely non-null component ${\phi_1}$ and the electromagnetic radiation ${\phi_2}$ (${\phi_2\ne0}$ requires ${\alpha\ne0}$). However, let us remark that, due to the freedom in the choice of the local null triad, under which the Maxwell scalars transform as (\ref{NPdefprimed}), at a given point there exists a special triad in which ${\phi'_2=0}$.

There is a \textbf{special case} ${Q=\hbox{const.}}$, for which the metric (\ref{RTmetric-final-factorized}) simplifies to
\begin{align}
\dd s^2 =&\ r^2 \big( \dd \varphi + \alpha(u)\,\dd u \big)^2 - 2\,\dd u\dd r  +\Big(m - \kappa_0\,Q^2 \ln \Big|\frac{Q}{r}\Big| + \Lambda\,r^2\Big)\, \dd u^2 \,, \label{RTmetric-final-Qconst}
\end{align}
where the rescaled constant reads ${m\equiv Q^2\mu}$, and the new coordinate is
\begin{equation}
\varphi = \frac{1}{Q} \int \frac{\dd x}{\beta(x)} \,.
\label{RTy-x}
\end{equation}

For ${\alpha(u)=0}$ (that is without the electromagnetic radiation component), and for compact coordinate $\varphi$, this family of spacetimes represents \emph{charged black holes} with any value of the cosmological constant~$\Lambda$. Indeed, by introducing the time coordinate~$t$ via the transfomation
\begin{equation}
\dd u = \dd t + \Big(m - \kappa_0\,Q^2 \ln \Big|\frac{Q}{r}\Big| + \Lambda\,r^2\Big)^{-1} \dd r\,,
\label{du->dt}
\end{equation}
we obtain the metric
\begin{align}
\dd s^2 =&
-\!\Big(\!-m + \kappa_0\,Q^2 \ln \Big|\frac{Q}{r}\Big| - \Lambda\,r^2\Big)\, \dd t^2
+\frac{\dd r^2}{\displaystyle -m + \kappa_0\,Q^2 \ln \Big|\frac{Q}{r}\Big| - \Lambda\,r^2}
+ r^2 \dd\varphi^2 \,, \label{RTmetric-tr-Qconst}
\end{align}
with the electromagnetic field
\begin{equation}
\bF =  \frac{Q}{r}\, \dd r \wedge \dd t
\qquad \hbox{corresponding to} \qquad
\bA =  Q \,\ln\frac{r}{r_0}\,  \dd t \,.
\label{RTformF-tr-Qconst}
\end{equation}
This is the standard form of \emph{cyclic symmetric, electrostatic solution with~$\Lambda$} in polar ``Schwarzschild'' coordinates found by Peldan in 1993 \cite{Peldan:1993}, see Eq.~(11.56) in~\cite{GarciaDiaz:2017}, which extended previous solutions by Gott, Simon and Alpert \cite{GottAlpert:1984,GottSimonAlpert:1986}, Deser and Mazur \cite{DeserMazur:1985}, and Melvin \cite{Melvin:1986} to any cosmological constant, see also Garc\'\i a \cite{Garcia:2009}. A thorough review and discussion of this class of solutions is contained in \cite{GarciaDiaz:2013} and also Section~11.2 of~\cite{GarciaDiaz:2017}.

For ${\alpha(u)\ne0}$ the spacetime (\ref{RTmetric-final-Qconst}) in general contains \emph{additional} electromagnetic \emph{radiation component} ${\phi_2\ne0}$. It remains to be analyzed in detail if such a situation can be physically interpreted as a charged black hole with a specific radiation, or if the function $\alpha(u)$ is just some kind of a kinematic parameter.

Similarly, the general Robinson--Trautman solution (\ref{RTmetric-final}) with aligned electromagnetic field (\ref{RTformF-final}) needs to be understood and explicitly related to other known solutions summarized in Chapter~11 of~\cite{GarciaDiaz:2017}, in particular the non-static ones. This seems to be in principle possible because, e.g., for ${e \ne 0}$ the transformation (\ref{du->dt}) introduces the metric component $g_{tx}$ typical for stationary spacetimes.

\newpage


\section{All non-aligned Robinson--Trautman solutions}
\label{sec:RTnonaligned}

After completing the systematic derivation of all aligned electromagnetic fields in the family of expanding Robinson--Trautman geometries, we now investigate the \emph{possible non-aligned fields}.

The Einstein--Maxwell equations are (\ref{RTEinstinEq}), in which  the functions $F_a$ are defined by (\ref{Fr})--(\ref{Fu}). The generic non-aligned electromagnetic field has
${\phi_0 \ne 0\ \Leftrightarrow \  F_{rx} \ne 0  \ \Leftrightarrow \ F_r \ne 0 }$.

\subsection{Integration of ${R_{rr}=\kappa_0\,G^2 F_r^2}$}
Using Eq.~(\ref{Ricci rr}) for the Ricci tensor component $R_{rr}$, we obtain the constraint
\begin{equation}
\kappa _0\, F_r^2  = -g_{xx}\,(\Theta_{,r}+\Theta^2) \,, \label{feqrrNE}
\end{equation}
where ${\Theta \ne 0}$ is the optical scalar representing the \emph{expansion} of the privileged null congruence generated by ${\mathbf{k}=\partial_r}$. Let us recall that it is directly related to the spatial metric function $g_{xx}$ via the relations
\begin{equation}
g_{xx}=G^{\,-2}\qquad\hbox{with}\qquad
\Theta = - (\ln G)_{,r} \equiv - \frac{G_{,r}}{G} \,,
\label{GRT}
\end{equation}
see (\ref{G}), (\ref{Theta-G}). Therefore, the metric component $g_{xx}$ must necessarily depend on the coordinate~$r$, otherwise ${\Theta=0}$.

It is possible to substitute from (\ref{GRT}) into (\ref{feqrrNE}), but we found more convenient to keep the expansion scalar $\Theta$ in (\ref{feqrrNE}). This equation \emph{explicitly expresses the non-aligned Maxwell field component ${F_{rx} \equiv F_r}$ in terms of the metric component $g_{xx}$} (and its $r$-derivatives via $G$). This relation can be rewritten as
\begin{equation}
\kappa _0\, F_{rx}^2  = G^{\,-2}\,\Theta^2 \big( (\Theta^{-1})_{,r} - 1 \big) \,. \label{feqrrNEalter}
\end{equation}

Notice that (in the Robinson--Trautman family) ${F_{rx} = 0  \ \Leftrightarrow \ \Theta^{-1}=r+r_{0}(u,x)}$. This fully corresponds to the previously studied aligned case, for which (\ref{ExplEx}) applies.

\subsection{Integration of ${R_{rx}=\kappa_0\,G^2 F_r F_x}$}
Using Eq.~(\ref{Ricci rp}) for the Ricci tensor component $R_{rx}$ and (\ref{feqrrNE}), we get the relation
\begin{equation}
\tfrac{1}{2}(\Theta\, g_{ux,r}-g_{ux,rr})
 =\kappa_0\,G^2 F_r \,( F_x  + g_{ux} F_r )  \,. \label{feqrxNE}
\end{equation}
In view of (\ref{Fr}), (\ref{Fx}), this is equivalent to
\begin{equation}
\kappa _0\, F_{ru} \, F_{rx}  = \tfrac{1}{2}(\Theta\, g_{ux,r}-g_{ux,rr}) \,. \label{feqrxNEalter}
\end{equation}
Therefore, \emph{by prescribing any metric function $g_{ux}$, the electromagnetic field component $F_{ru}$ is explicitly determined}.

Notice that it \emph{admits a special solution} ${F_{ru}=0 \ \Leftrightarrow \ \Theta\, g_{ux,r}=g_{ux,rr}}$. This occurs \emph{either} when ${g_{ux}}$ is independent of the coordinate~$r$,
\begin{equation}
g_{ux}= B(u,x) \,, \label{RTnonaligned-f}
\end{equation}
\emph{or}, using (\ref{GRT}), when ${\Theta = (\ln G^{\,-1})_{,r} = (\ln g_{ux,r})_{,r}}$
which can be completely integrated as
\begin{equation}
g_{xx} = A(u,x) \,\big(g_{ux,r}\big)^2\,,
\label{the case Fru=0 inRTintegr}
\end{equation}
where ${A>0}$ is any function independent of~$r$.

\subsection{Integration of ${R_{ru}=-2\Lambda + \kappa_0\,G^2 F_r F_u}$}
The generic Ricci tensor component $R_{ru}$ is given by (\ref{Ricci ru}), so that the corresponding Einstein--Maxwell field equation becomes
\begin{align}
-\tfrac{1}{2}g_{uu,rr}+\tfrac{1}{2}g^{rx}g_{ux,rr}+\tfrac{1}{2}g^{xx}\big(g_{ux,r||x}+(g_{ux,r})^2\big)& \nonumber \\
-\Theta_{,u}-\tfrac{1}{2}\Theta\big(g^{xx}g_{xx,u}+g^{rx}g_{ux,r}+g_{uu,r}\big)
&= -2\Lambda + \kappa_0\,G^2  F_r \,F_u \,. \label{feqruNE}
\end{align}
This uniquely \emph{determines the third electromagnetic field component} (\ref{Fu}) represented by $F_u$.
Using (\ref{Fr})--(\ref{Fu}) and then (\ref{feqrrNE}), (\ref{feqrxNEalter}), the last term on the right-hand side can be expressed as
\begin{align}
\kappa _0\, G^2 &\, F_{rx}\,(g_{ux}F_{ru}-F_{ux}-g_{uu}F_{rx}) \nonumber \\
&= \kappa _0\, g^{rx}F_{ru}\,F_{rx}\,-\kappa _0\, g^{xx}F_{ux}\,F_{rx}\,-\kappa _0\, g^{xx}g_{uu} \,F_{rx}^2 \nonumber \\
&= -\kappa _0\, g^{xx}F_{ux}\,F_{rx}
+\tfrac{1}{2} g^{rx}(\Theta\, g_{ux,r}-g_{ux,rr})
+ g_{uu} \big(\Theta_{,r}+\Theta^2\big)   \,. \label{feqruNE-FrFu}
\end{align}
The field equation (\ref{feqruNE}) thus reads
\begin{align}
\kappa _0\, F_{ux}\,F_{rx}
=&\
\tfrac{1}{2}g_{xx} \big( g_{uu,rr} + \Theta\, g_{uu,r} + 2(\Theta_{,r}+\Theta^2) g_{uu} -4\Lambda \big) \label{feqruNEalter}\\
&
+g_{ux}(\Theta\, g_{ux,r}-g_{ux,rr})
-\tfrac{1}{2}\big(g_{ux,r||x}+(g_{ux,r})^2\big)
+\tfrac{1}{2}\Theta\,g_{xx,u}
+g_{xx}\Theta_{,u}
\nonumber \,.
\end{align}
\emph{By prescribing any metric function $g_{uu}$, the third electromagnetic field component $F_{ux}$ is thus explicitly determined}.

To summarize, by employing three (out of six) independent components of the Einstein field equations, we have now derived explicit expressions (\ref{feqrrNEalter}), (\ref{feqrxNEalter}) and (\ref{feqruNEalter}) which \emph{determine all three components of the electromagnetic field}, namely $F_{rx}$, $F_{ru}$ and $F_{ux}$, respectively, in terms of the three (so far) independent metric components $g_{xx}$, $g_{ux}$ and $g_{uu}$.

These three expressions are equivalent to equations (\ref{feqrrNE}), (\ref{feqrxNE}), (\ref{feqruNE}) for the three dual electromagnetic functions ${F_a \equiv \,^*\!F_{a}/G}$. They can be written in a very compact form
\begin{align}
\kappa_0\, F_r^2  &   = \alpha\,, \label{FrEq-NE}\\
\kappa_0\, F_r\,F_x & = \beta - \alpha\,g_{ux}\,, \label{FxEq-NE}\\
\kappa_0\, F_r\,F_u & = \gamma\,, \label{FuEq-NE}
\end{align}
where the functions $\alpha, \beta, \gamma$ are useful shorthands for the combination of the three metric functions
\begin{align}
\alpha \equiv &-g_{xx}\,(\Theta_{,r}+\Theta^2) \,, \label{alfa}\\
\beta  \equiv &\ \tfrac{1}{2} \,g_{xx}\,(\Theta\, g_{ux,r}-g_{ux,rr}) \,, \label{beta}\\
\gamma \equiv &\ \tfrac{1}{2} \big[\, g_{xx}(4\Lambda-g_{uu,rr})
+g_{ux}\,g_{ux,rr}+g_{ux,r||x}+(g_{ux,r})^2& \label{gama}\\
 & \hspace{5mm} -2g_{xx}\Theta_{,u}-\Theta\,(g_{xx,u}+g_{ux}\,g_{ux,r}+g_{xx}\,g_{uu,r})\big]\,. \nonumber
\end{align}
Consequently,
\begin{align}
F_r = \sqrt{\frac{\alpha}{\kappa_0}}\,, \qquad
F_x = \Big(\frac{\beta}{\alpha} - \,g_{ux}\Big) F_r\,, \qquad
F_u = \frac{\gamma}{\alpha}\,F_r\,. \label{Fx/Fr-Fu/Fr}
\end{align}

Let us recall that $\alpha$ is fully determined by $g_{xx}$, the function $\beta$ is determined by $g_{xx}$ and $g_{ux}$, while the third metric component $g_{uu}$ enters only $\gamma$.

\newpage

\subsection{The Maxwell equations}
As the next step, we apply the 4 independent Maxwell equations in the form  (\ref{Maxeq-dual}) and (\ref{Maxeq2}), namely
\begin{align}
(G\,F_{a})_{,b} = \big(G\,F_{b})_{,a}
\qquad\hbox{and}\qquad
F_{ux,r}+F_{ru,x}-F_{rx,u} = 0\,,
\label{Maxeq-NE}
\end{align}
which restrict the possible electromagnetic field and its coupling to the gravitational field. For explicit evaluation of the partial derivatives with respect to ${a,b=\{r,u,x \}}$ we employ the expressions directly following from (\ref{GRT}) and (\ref{FrEq-NE})--(\ref{FuEq-NE}), implying (\ref{Fx/Fr-Fu/Fr}), namely
\begin{align}
G_{,a}  & = -\tfrac{1}{2}G^3\, g_{xx,a}\,, \label{derG}\\
F_{r,a} & = \frac{1}{\kappa_0 F_r}\,\Big(\tfrac{1}{2}\alpha_{,a}\Big), \label{derFrEq-NE}\\
F_{x,a} & = \frac{1}{\kappa_0 F_r}\,\Big((\beta - \alpha\,g_{ux})_{,a} -
\tfrac{1}{2}(\beta - \alpha\,g_{ux})\frac{\alpha_{,a}}{\alpha} \Big), \label{derFxEq-NE}\\
F_{u,a} & = \frac{1}{\kappa_0 F_r}\,\Big(\gamma_{,a} - \tfrac{1}{2}\gamma\frac{\alpha_{,a}}{\alpha} \Big). \label{derFuEq-NE}
\end{align}
Using these relations in calculating ${(G\,F_{a})_{,b} = \big(G\,F_{b})_{,a}}$ for ${ab=rx, ru, ux}$ we obtain
\begin{align}
\Big(\alpha_{,x} + 2\alpha\,\frac{G_{,x}}{G}\Big) - 2(\beta - \alpha\,g_{ux})_{,r} + (\beta - \alpha\,g_{ux})\Big( \frac{\alpha_{,r}}{\alpha}  + 2\Theta\Big) & = 0 \,, \label{Max1}\\
\Big(\alpha_{,u} + 2\alpha\,\frac{G_{,u}}{G}\Big) - 2\gamma_{,r} + \gamma\,\Big( \frac{\alpha_{,r}}{\alpha} + 2\Theta \Big) & = 0 \,, \label{Max2}\\
\gamma_{,x} - \gamma\,\Big( \frac{\alpha_{,x}}{2\alpha} - \frac{G_{,x}}{G} \Big)
- (\beta - \alpha\,g_{ux})_{,u}
+ (\beta - \alpha\,g_{ux})\Big( \frac{\alpha_{,u}}{2\alpha} - \frac{G_{,u}}{G}\Big) & = 0 \,, \label{Max3}
\end{align}
respectively. Notice that the terms in the large brackets depend only on ${g_{xx}\equiv G^{\,-2}}$ and their derivatives. The last Maxwell equation (\ref{Maxeq-NE}), using the inversion of (\ref{Fr})--(\ref{Fu}),
\begin{align}
F_{rx} & = F_r\,, \label{Frinv}\\
F_{ru} & = G^2 (F_x + g_{ux}F_r)\,, \label{Fxinv}\\
F_{ux} & = g_{ux}\,G^2 (F_x + g_{ux}F_r) - F_u - g_{uu}F_r\,, \label{Fuinv}
\end{align}
reads
\begin{align}
\beta_{,x} + \beta_{,r}\,g_{ux} + \beta\,\Big[\, g_{ux,r} - g_{ux}\,\Big( \frac{\alpha_{,r}}{2\alpha} + 2 \Theta \Big) - \Big( \frac{\alpha_{,x}}{2\alpha} - 2\frac{G_{,x}}{G} \Big)\Big] & \nonumber\\
- \frac{1}{2G^2} \Big[\, 2\gamma_{,r} - \alpha_{,r}\,\Big( \frac{\gamma}{\alpha} - g_{uu} \Big)
+ \alpha_{,u} + 2\alpha\,g_{uu,r} \Big] & = 0 \,. \label{Max4}
\end{align}
The four equations (\ref{Max1})--(\ref{Max3}) and (\ref{Max4}) put restrictions on the metric functions, encoded in $G, \alpha, \beta, \gamma$.

\subsection{Remaining Einstein  equations ${R_{ab} = 2\Lambda\,g_{ab} + \kappa _0\,G^2 F_a F_b}$}
Finally, it is necessary to solve the remaining three Einstein equations (\ref{EinstinEq}) for the components ${ab=xx, ux, uu}$. Using (\ref{Fx/Fr-Fu/Fr}) we immediately derive their form
\begin{align}
R_{xx} & = 2\Lambda\,g_{xx} + \frac{G^2}{\alpha}\,(\beta - \alpha\,g_{ux})^2\,, \label{xxEq-NE}\\[1mm]
R_{ux} & = 2\Lambda\,g_{ux} + \frac{G^2}{\alpha}\,(\beta - \alpha\,g_{ux})\,\gamma\,, \label{uxEq-NE}\\[1mm]
R_{uu} & = 2\Lambda\,g_{uu} + \frac{G^2}{\alpha}\,\gamma^2\,. \label{uuEq-NE}
\end{align}
Substituting the explicit expressions for the corresponding Ricci tensor components (\ref{Ricci pq})--(\ref{Ricci uu}) reveals a rather complicated system of PDEs for the metric functions which must be solved together with (\ref{Max1})--(\ref{Max3}) and (\ref{Max4}).

At this stage, it does not seem possible to find a \emph{general} solution of these equations. However, \emph{we have achieved a separation of the variables representing the gravitational and the electromagnetic field}. Indeed, the system of 7 equations (\ref{Max1})--(\ref{Max3}), (\ref{Max4}), and (\ref{xxEq-NE})--(\ref{uuEq-NE}) with (\ref{Ricci pq})--(\ref{Ricci uu}) \emph{involves only the three metric functions} $g_{xx}$, $g_{ux}$, $g_{uu}$, encoded also in the functions $G$ and $\alpha, \beta, \gamma$ defined in (\ref{GRT}) and (\ref{alfa})--(\ref{gama}). After their solution is found, the corresponding three (dual) components of the electromagnetic field $F_{r}$, $F_{x}$, $F_{u}$ are easily obtained by applying the relations (\ref{Fx/Fr-Fu/Fr}). The components $F_{rx}$, $F_{ru}$, $F_{ux}$ are then their simple combinations (\ref{Frinv})--(\ref{Fuinv}).

\subsection{A simple particular solution}
\label{particular-solution-RT}
To demonstrate the usefulness of our formulation of the most general Einstein--Maxwell field equations and also \emph{to show that the class of Robinson--Trautman 2+1 spacetimes with non-aligned electromagnetic field is not empty}, we will now derive a special solution of the above system of equations.

Let us \emph{assume} that \emph{only the non-aligned component} $F_r$ of the electromagnetic field is non-trivial, i.e.,
\begin{align}
F_r = \sqrt{\frac{\alpha}{\kappa_0}} \ne 0\,, \qquad
F_x = 0\,, \qquad
F_u = 0\,. \label{simpleF}
\end{align}
The field equations (\ref{FrEq-NE})--(\ref{FuEq-NE}) then imply
\begin{align}
\beta - \alpha\,g_{ux} & = 0 \,, \label{Fx=0=Fu-1}\\
\gamma & = 0\,. \label{Fx=0=Fu-2}
\end{align}
Further simplification is achieved by \emph{assuming}
\begin{align}
g_{ux} = 0\,. \label{gux=0}
\end{align}
In such a case the condition (\ref{Fx=0=Fu-1}) ${\beta=0}$ is satisfied due to (\ref{beta}), while (\ref{Fx=0=Fu-2}) gives
\begin{align}
g_{uu,rr} - 4\Lambda + 2\Theta_{,u} + \Theta \Big( g_{uu,r} - 2\frac{G_{,u}}{G} \Big) & = 0\,.  \label{gama=0}
\end{align}
The Maxwell equations (\ref{Max1})--(\ref{Max3}), (\ref{Max4}) reduce to
\begin{align}
\frac{\alpha_{,x}}{\alpha} + 2\,\frac{G_{,x}}{G}  & = 0 \,, \label{Max1-Fx=0=Fu}\\
\frac{\alpha_{,u}}{\alpha} + 2\,\frac{G_{,u}}{G}  & = 0 \,, \label{Max2-Fx=0=Fu}\\
\alpha_{,r}\, g_{uu} + \alpha_{,u} + 2\alpha\,g_{uu,r} & = 0 \,, \label{Max4-Fx=0=Fu}
\end{align}
and the final three Einstein equations simplify as
\begin{align}
R_{xx} & = 2\Lambda\,g_{xx} \,, \label{xxEq-NE-Fx=0=Fu}\\[1mm]
R_{ux} & = 0 \,, \label{uxEq-NE-Fx=0=Fu}\\[1mm]
R_{uu} & = 2\Lambda\,g_{uu} \,, \label{uuEq-NE-Fx=0=Fu}
\end{align}
where
\begin{align}
R_{xx} & = g_{xx}\,g_{uu}\big(\Theta_{,r} +\Theta^2 \big) +2g_{xx}\Theta_{,u}
     +\Theta\big(g_{xx}g_{uu,r} + g_{xx,u}\big) \,, \label{Ricci pq-spec} \\
R_{ux} & = -\tfrac{1}{2}g_{uu,xr} +\tfrac{1}{2}\Theta g_{uu,x}  \,, \label{Ricci up-spec} \\
R_{uu} & = \tfrac{1}{2}g_{uu}g_{uu,rr} + \tfrac{1}{4}g^{xx} g_{xx,u} g_{uu,r}-\tfrac{1}{2}g^{xx}g_{xx,uu} \nonumber \\
 & \hspace{5.0mm} -\tfrac{1}{2}g^{xx}g_{uu||xx}+\tfrac{1}{4}(g^{xx} g_{xx,u})^2
+\tfrac{1}{2}\Theta\big(g_{uu}g_{uu,r}-g_{uu,u}\big) \,. \label{Ricci uu-spec}
\end{align}

The equations (\ref{Max1-Fx=0=Fu}) and (\ref{Max2-Fx=0=Fu}) can be easily integrated, yielding
\begin{align}
\alpha = f(r)\,G^{-2} \equiv f(r)\,g_{xx}\,, \label{alpha=g/G2}
\end{align}
where $f(r)$ is any function of the coordinate $r$. Equation (\ref{Max4-Fx=0=Fu}) gives the constraint
\begin{align}
g_{uu,r} + \Big(\frac{f'}{2f}+\Theta\Big) g_{uu} - \frac{G_{,u}}{G}  & = 0 \,, \label{Max4-next}
\end{align}
in which $f'$ is the derivative of $f$. It thus remains to solve (\ref{gama=0}), (\ref{Max4-next}) and
(\ref{xxEq-NE-Fx=0=Fu})--(\ref{uuEq-NE-Fx=0=Fu}).

Now, combining (\ref{alpha=g/G2}) with the definition (\ref{alfa}) we obtain
$$ \Theta_{,r}+\Theta^2 = - f(r) \,, $$
which is the Ricatti-type equation for the expansion $\Theta$.
Using the substitution ${\Theta = z_{,r}/z}$, it can be rewritten as the linear equation ${z_{,rr}+ f(r)\,z =0}$. Let us consider here only \emph{the simplest case of a constant}~${f}$,
\begin{align}
f\equiv C^2\,. \label{f=C^2}
\end{align}
By applying (\ref{GRT}) we obtain the explicit solution
\begin{align}
\Theta(r) & = C\,\cot(Cr) \,, \label{Theta-C}\\[1mm]
G & = \frac{P(u,x)}{\sin(Cr)} \,, \label{G-C} \\[1mm]
g_{xx} & = \frac{\sin^2(Cr)}{P^2(u,x)} \,. \label{gxx-C}
\end{align}
(We have applied the coordinate freedom, namely a trivial constant shift in the coordinate $r$, to simplify the expressions.) It is now easily seen that for the particular choice
\begin{align}
P & = 1 \,, \label{P-C}\\
g_{uu}  & = 0 \,, \label{guu-C} \\
\Lambda & = 0\,, \label{Lambda-C}
\end{align}
all the remaining field equations (\ref{gama=0}), (\ref{Max4-next}) and
(\ref{xxEq-NE-Fx=0=Fu})--(\ref{uuEq-NE-Fx=0=Fu}) are satisfied because
${R_{xx} = 0}$, ${R_{ux} = 0}$ and ${R_{uu} = 0}$. We have thus obtained a \emph{special Robinson--Trautman solution}
\begin{equation}
\dd s^2 = \sin^2(Cr)\, \dd x^2 - 2\,\dd u\,\dd r \,, \label{nonalRT}
\end{equation}
\emph{with a non-aligned electromagnetic field}
\begin{align}
F_r = \frac{C}{\sqrt{\kappa_0}\,G} = \frac{C}{\sqrt{\kappa_0}}\,\sin(Cr)\,, \qquad
F_x = 0\,, \qquad
F_u = 0\,, \label{simplestF}
\end{align}
that is
\begin{align}
^*\bF = \frac{C}{\sqrt{\kappa_0}}\, \dd r\,.
\label{*Fasimpelst}
\end{align}
Using (\ref{Frinv})--(\ref{Fuinv}), this is equivalent to
\begin{equation}
\bF = \frac{C}{\sqrt{\kappa_0}}\,\sin(Cr)\, \dd r \wedge \dd x  \,,
\label{2-formFsimplest}
\end{equation}
corresponding to the potential
\begin{equation}
\bA = -\frac{1}{\sqrt{\kappa_0}}\,\cos(Cr)\, \dd x  \,.
\label{RTformA-simplest}
\end{equation}
By rescaling the coordinates $r$ and $u$ the constant $C$ can be set to ${C=1}$, but we prefer to keep it free because it represents the value of the electromagnetic field and $r$ is not dimensionless.

Actually, (\ref{nonalRT}) is the metric 3) on page 133 of \cite{Kruchkovich:1955} for ${q=0}$, which admits 4 Killing vectors (see also the metric (4.1) in \cite{Chow:2019}).

\newpage


\section{Final summary and remarks}
\label{sec:finalsummary}

In this contribution we systematically solved the Einstein--Maxwell equations with $\Lambda$, obtaining all electrovacuum 2+1 spacetimes. We identified main geometrically distinct subclasses, and we explicitly derived the corresponding metrics and electromagnetic fields. In particular:

\begin{itemize}

\item
The metric of \emph{any} such spacetime can be written in canonical coordinates
in the form (\ref{general nontwist})
\begin{equation}
\dd s^2 = \,G^{-2}\,\dd x^2+2\,g_{ux}\, \dd u\, \dd x -2\,\dd u\,\dd r + g_{uu}\, \dd u^2 \,.
\label{general nontwist_summary}
\end{equation}

\item
The \emph{generic} electromagnetic Maxwell 2-form field and its dual 1-form have three independent components (\ref{2-formF}) and (\ref{*Fa}), namely
\begin{align}
\bF   &= F_{ru}\, \dd r \wedge \dd u + F_{rx}\, \dd r \wedge \dd x  + F_{ux}\, \dd u \wedge \dd x\,,\\
^*\bF &= G\,(\, F_{r}\, \dd r + F_{u}\, \dd u + F_{x}\, \dd x \,)\,,
\label{2-formF_summary}
\end{align}
where
${\,F_r = F_{rx}}$,
${\,F_x = g_{xx}F_{ru}-g_{ux}F_{rx}}$,
${\,F_u = g_{ux}F_{ru}-F_{ux}-g_{uu}F_{rx}}$.

\item
In terms of the Newman--Penrose scalars (\ref{NPdef}) of distinct boost weights $+1$, $0$, ${-1}$, the Maxwell field \emph{invariants} ${F^2 \equiv F_{ab}\,F^{ab}}$ and ${\,^*\!F^2 \equiv \,^*\!F_{a} \,^*\!F^{a}}$ are
\begin{equation}
\tfrac{1}{2}F^2 = -\,^*\!F^2 = 2\phi_0\phi_2-\phi_1^2\,.
\label{F2invariant-scalars_summary}
\end{equation}
The electromagnetic field is \emph{aligned with\,}
${\mathbf{k}=\partial_r \Leftrightarrow \phi _0 = 0  \Leftrightarrow F_{rx} = 0  \Leftrightarrow F_r =0}$.

Such an aligned field has only two components, namely ${\phi _2 = G\, F_u\equiv G\,( g_{ux}F_{ru}-F_{ux})}$ and ${\phi _1 = G^2 F_x \equiv F_{ru}}$. In the case when ${\phi _2 = 0 \Leftrightarrow  F_u=0}$, the electromagnetic field is \emph{non-null}, characterized just by ${\phi _1=F_{ru}}$. Contrarily, when ${\phi _1 = 0 \Leftrightarrow  F_x=0}$, it is \emph{null (radiative)}, characterized just by ${\phi _2 = -G\,F_{ux}}$.

\item
Evaluating the energy-momentum tensor (\ref{Tab}) we derived that, in terms of these quantities, the \emph{Einstein--Maxwell field equations take a simple form} (\ref{EinstinEq}),
\begin{equation}
R_{ab} = 2\Lambda\,g_{ab} + \kappa _0\,G^2 F_a F_b \,, \label{EinstinEq_summary}
\end{equation}
(equivalent to ${R_{ab} = 2\Lambda\,g_{ab} + \kappa_0 \,^*\!F_{a}\!\,^*\!F_{b} }$) and (\ref{Maxeq-dual}), (\ref{Maxeq2}),
\begin{align}
(G\,F_{a})_{,b} = \big(G\,F_{b})_{,a}\,, \qquad
 F_{[ab,c]} =0\,. \label{Maxeq12}
\end{align}

\item
In the triad (\ref{triadexpl}) of the metric (\ref{general nontwist_summary}), all \emph{optical scalars} of a congruence generated by the privileged null vector field~${\mathbf{k}=\partial_r}$ \emph{vanish except}, possibly,  \emph{expansion}
\begin{equation}
\Theta = - (\ln G)_{,r} \,.
\label{Theta-G_summary}
\end{equation}
There are thus \emph{two geometrically distinct classes of spacetimes} to be investigated:

\begin{enumerate}

\item ${\Theta=0}$, defining the non-expanding \emph{Kundt class}, with the metric function
\begin{equation}
G \equiv P(u,x)\,, \label{GKundt_summary}
\end{equation}

\item ${\Theta\neq 0}$, defining the expanding \emph{Robinson--Trautman class}, with the metric function
\begin{equation}
G \equiv G(r,u,x)\,. \label{GRT_summary}
\end{equation}

\end{enumerate}

\item
Keeping the full generality, we explicitly integrated the coupled system of the field equations (\ref{EinstinEq_summary}) and (\ref{Maxeq12}) \emph{both} for the Kundt and the Robinson--Trautman spacetimes. It turned out that, as in standard 3+1 general relativity, the \emph{Kundt class only admits aligned electromagnetic fields} while the \emph{Robinson--Trautman class admits both aligned and non-aligned electromagnetic fields}. Therefore, we treated these three distinct families of spacetimes in three separate sections of our paper, namely  Sec.~\ref{sec:Kundt}, Sec.~\ref{sec:RTaligned}, and Sec.~\ref{sec:RTnonaligned}, respectively.

\item
All \emph{Kundt} spacetimes (Sec.~\ref{sec:Kundt}) with  \emph{necessarily aligned electromagnetic fields} have the form
\begin{align}
\dd s^2 =&\
\frac{\dd x^2}{P^2} + 2\,(e+f\,r)\, \dd u\, \dd x -2\,\dd u\,\dd r \nonumber \\
& +\Big(a+b\,r+\big( \Lambda + \tfrac{1}{4} P^2f^2 - \tfrac{ \kappa _0 }{2} Q^2 \big)\,r^2\Big)\, \dd u^2 \,, \label{Kmetric-final_summary}
\end{align}
and
\begin{equation}
\bF =  Q\, \dd r \wedge \dd u + (f\,Q\,r-\xi)\, \dd u \wedge \dd x \,,
\label{KformF-final_summary}
\end{equation}

corresponding to the potential
\begin{equation}
\bA = A_r\,\dd r + A_x\,\dd x \,,
\label{KformA-summary}
\end{equation}
where ${A_r = -\int \! Q\, \dd u\,}$ and ${A_x = r \int \! f\,Q\, \dd u  - \int \! \xi\, \dd u}$,
see equations~(\ref{Kmetric-final})--(\ref{KformA-final}). As summarized in Subsec.~\ref{summary-Kundt}, the function $Q(u,x)$ represents the \emph{non-null} component, while the function $\xi(u,x)$ represents the \emph{null} component of the Maxwell field. Their relation to the metric functions $P, e, f$ and $a, b$ is explicitly given by the Einstein--Maxwell equations (\ref{KFconstraint-summary1})--(\ref{KFconstraint-summary5}). In Subsec.~\ref{summary-Kundt} we presented a basic description of these solutions, separately for \emph{two geometrically distinct subclasses} ${f=0}$  and ${f\ne0}$.

This large family of non-expanding Kundt spacetimes contains many interesting subclasses which represent electrovacuum universes and also waves on these cosmological backgrounds. The simplest of them are gravitational and electromagnetic \emph{pp-waves} with ${\Lambda=0}$. These are defined by the condition ${k_{a;b}=\frac{1}{2}g_{ab,r}=0}$ which requires ${f=0}$, ${b=0}$, ${Q = 0}$. The field equations (\ref{KFconstraint-summary1-f=0})--(\ref{KFconstraint-summary5f=0}) then yield the explicit metric in the Brinkmann form \cite{Brinkmann:1925}
\begin{eqnarray}
\dd s^2 \rovno \frac{\dd x^2}{P^2} +2\,e\,\dd u \dd x -2\,\dd u\dd r  + a\, \dd u^2 \,, \label{pp-metric}
\end{eqnarray}
and the coupled electromagnetic wave
\begin{equation}
\bF = -\frac{\gamma(u)}{P(u,x)}\, \dd u \wedge \dd x \,,
\label{KformF-final-pp}
\end{equation}
corresponding to
\begin{equation}
\bA = A_x\,\dd x \quad \hbox{where}\quad  A_x = - \int \! \frac{\gamma(u)}{P(u,x)}\, \dd u \,.
\label{KformF-final-pp+}
\end{equation}
Here $\gamma(u)$ is an \emph{arbitrary profile function} of the retarded time~$u$, while the metric function $a(u,x)$ is obtained by integrating the only remaining field equation (\ref{KFconstraint-summary5f=0}).

\item
All \emph{Robinson--Trautman} spacetimes (Sec.~\ref{sec:RTaligned}) with  \emph{aligned electromagnetic fields} (for which the metric function $G$ simplifies to ${G=P(u,x)/r}$) can be written as
\begin{align}
\dd s^2 =&\  \frac{r^2}{P^2}\,
\big( \dd x + e \,P^2 \dd u \big)^2 -2\,\dd u\dd r \nonumber \\
& +\Big(\mu\, Q^2 - \kappa_0\,Q^2 \ln \Big|\frac{Q}{r}\Big|
  +2\,(\ln Q)_{,u}\,r + \Lambda\,r^2\Big)\, \dd u^2  \label{RTmetric-final_summary}
\end{align}
with
\begin{equation}
\bF =  \frac{Q(u)}{r}\, \dd r \wedge \dd u
\qquad \hbox{corresponding to} \qquad
\bA =  Q(u) \,\ln\frac{r}{r_0}\,  \dd u \,,
\label{RTformF-final_summary}
\end{equation}
see equations~(\ref{RTmetric-final})--(\ref{RTformA-final}). Here $\mu$ is a constant while the metric functions $P$ and $e$ satisfy the field equation (\ref{RTFconstraint-2Q}), that is
\begin{equation}\label{RTFconstraint-2Q_summary}
\Big(\frac{Q}{P}\Big)_{,u} = Q\,(e\,P)_{,x} \,.
\end{equation}
The dual 1-form Maxwell field reads
\begin{equation}
^*\bF = \frac{Q}{P}\, \dd x + e\,P\,Q\, \dd u  \,.
\label{RTformF-final_sum}
\end{equation}

As summarized in Subsec.~\ref{summary-RT}, the function $Q(u)$ gives the \emph{non-null} component ${\phi _1 = Q(u)/r}$ of the Maxwell field. Somewhat surprisingly, there is also an additional \emph{null} (radiative) component ${\phi_2 = e\,P\,Q}$ when ${e\ne0}$. However, such Maxwell fields \emph{cannot be purely null} because ${\phi _1 = 0}$ implies ${\phi _2 = 0}$.

The simplest ${e \ne 0}$ solution of the field equation (\ref{RTFconstraint-2Q_summary}) is
${P=1}$, ${e= x\,(\ln Q)_{,u} + \alpha(u)}$, which yields the metric (\ref{RTmetric-simplest}).

Another interesting subclass (\ref{RTmetric-final-factorized}) arises for factorized $P$ such that ${ P=Q(u)\,\beta(x)}$ and ${e\,P=\alpha(u)}$. The special case ${\alpha=0}$ and ${Q=\hbox{const.}}$ of these expanding Robinson--Trautman spacetimes is equivalent to the solution (\ref{RTmetric-tr-Qconst}), (\ref{RTformF-tr-Qconst}),
\begin{align}
\dd s^2 = -\Phi(r)\, \dd t^2 + \frac{\dd r^2}{\Phi(r)} + r^2 \dd\varphi^2 \,,
\qquad  \Phi(r)= -m + \kappa_0\,Q^2 \ln \Big|\frac{Q}{r}\Big| - \Lambda\,r^2\,,
\label{RTmetric-tr-Qconst_summary}
\end{align}
which is the family of \emph{cyclic symmetric, electrostatic black holes with~$\Lambda$} found in \cite{Peldan:1993} and discussed in Section~11.2 of~\cite{GarciaDiaz:2017}.

\item
The complementary class of \emph{Robinson--Trautman} spacetimes  with  \emph{non-aligned electromagnetic fields} is presented in Sec.~\ref{sec:RTnonaligned}. In this more complex case, the metric has the form (\ref{general nontwist_summary}) with a general function ${G(r,u,x)}$, cf.~(\ref{GRT_summary}). Moreover, the electromagnetic field now has a nontrivial component ${\phi_0 \ne 0\ \Leftrightarrow \  F_{rx} \ne 0  \ \Leftrightarrow \ F_r \ne 0 }$, which considerably complicates the solution of the Einstein--Maxwell equations.

Nevertheless, we were able to explicitly express the generic three components of the Maxwell field \emph{separately} in terms (of the combination) of the \emph{metric functions} as
\begin{align}
F_r = \sqrt{\frac{\alpha}{\kappa_0}}\,, \qquad
F_x = \Big(\frac{\beta}{\alpha} - \,g_{ux}\Big) F_r\,, \qquad
F_u = \frac{\gamma}{\alpha}\,F_r\,, \label{Fx/Fr-Fu/Fr_summary}
\end{align}
where the functions $\alpha, \beta, \gamma$ are defined in (\ref{alfa})--(\ref{gama}). Interestingly, $\alpha$ is determined only by~$g_{xx}$, $\beta$ is determined by $g_{xx}$ and $g_{ux}$, while the third metric component $g_{uu}$ enters only~$\gamma$.

We also derived a fully explicit form (\ref{Max1})--(\ref{Max3}), (\ref{Max4}) of all 4 Maxwell equations (\ref{Maxeq12}). Finally, there are 3 remaining Einstein equations (\ref{xxEq-NE})--(\ref{uuEq-NE}). This system of 7 equations involves only 3 \emph{metric} functions. After their solution is found, all components $F_{r}$, $F_{x}$, $F_{u}$ of the corresponding electromagnetic field are easily obtained using (\ref{Fx/Fr-Fu/Fr_summary}). In this sense, \emph{we have achieved a separation of the variables representing the gravitational and the electromagnetic field}.

Although at present it is not possible for us to find a \emph{general} solution to these 7 equations, the formulation of the problem presented here seems to be useful. This fact has been demonstrated in Subsec.~\ref{particular-solution-RT} where we have explicitly identified a  \emph{particular solution with non-aligned electromagnetic field}
\begin{equation}
\dd s^2 = \sin^2(Cr)\, \dd x^2 - 2\,\dd u\,\dd r \,, \label{nonalRT_summary}
\end{equation}
with
\begin{equation}
\bF = \frac{C}{\sqrt{\kappa_0}}\,\sin(Cr)\, \dd r \wedge \dd x
\qquad \hbox{corresponding to} \qquad
\bA = -\frac{1}{\sqrt{\kappa_0}}\,\cos(Cr)\, \dd x  \,.
\label{2-formFsimplest_summary}
\end{equation}
This special exact Robinson--Trautman spacetime contains electromagnetic field which has \emph{only} the non-aligned component ${F_r = (C/\!\sqrt{\kappa_0}\,)\,\sin(Cr)}$. It admits 4 Killing vectors \cite{Kruchkovich:1955,Clement:1993,Chow:2019}.

\end{itemize}

Of course, many questions have remained open. First of all, it is necessary to find explicit relations to already known solutions summarized in \cite{GarciaDiaz:2017}. Some basic identifications have already been presented here, namely:
\newpage
 
\begin{itemize}
\item Maximally symmetric backgrounds (Minkowski, de~Sitter, anti-de Sitter) are contained both in the Kundt and Robinson--Trautman class of spacetimes (\ref{Kmetric-final_summary}) and (\ref{RTmetric-final_summary}), respectively.
\item There are electrovacuum backgrounds in the form of direct-product geometries, such as the 2+1 analogue of the exceptional Pleba\'nski--Hacyan metric with ${\Lambda<0}$ and uniform Maxwell field (\ref{Kmetric-final-f=0-background}).
\item We identified the complete family of pp-waves in flat space, which are spacetimes admitting a covariantly constant null vecor field. In the Brikmann form (\ref{pp-metric}) they include the off-diagonal metric terms.
\item Within the Robinson--Trautman class with aligned fields we explicitly identified the cyclic symmetric charged black holes with any cosmological constant and electrostatic field (\ref{RTmetric-tr-Qconst_summary}).
\end{itemize}

Our main problem now is to identify all other known classes of solutions in 2+1 dimensions by using specific invariant geometrical characterizations (such as an algebraic structure, symmetries, identification of rotation and acceleration of the sources, etc.). Subsequently, explicit coordinate transformation must be found to relate our form of the solutions to those derived previously.

After identification of new spacetimes, their geometrical and physical analysis should be performed. Also, a systematic integration of the field equations for non-aligned Maxwell fields in the Robinson--Trautman class is desirable. However, these tasks are left for future works.

\section*{Acknowledgments}

This work was supported by the Czech Science Foundation Grant No.~GA\v{C}R 20-05421S. We are also grateful to the referee for many useful comments and suggestions which helped us to improve this contribution.

\newpage

\appendix

\section{Connections and curvature components in canonical coordinates}
\label{sec_geomcurv}
\renewcommand{\theequation}{A\arabic{equation}}
\setcounter{equation}{0}

The Christoffel symbols for the general non-twisting spacetime (\ref{general nontwist}) after applying the condition (\ref{shearfree condition}) are
\begin{eqnarray}
&& {\textstyle \Gamma^r_{rr} = 0} \,, \label{ChristoffelBegin} \\
&& {\textstyle \Gamma^r_{ru} = -\frac{1}{2}g_{uu,r}+\frac{1}{2}g^{rx}g_{ux,r}} \,, \\
&& {\textstyle \Gamma^r_{rx} = -\frac{1}{2}g_{ux,r}+\Theta g_{ux}} \,, \\
&& {\textstyle \Gamma^r_{uu} = \frac{1}{2}\big[-g^{rr}g_{uu,r}-g_{uu,u}+g^{rx}(2g_{ux,u}-g_{uu,x})\big]} \,, \\
&& {\textstyle \Gamma^r_{ux} = \frac{1}{2}\big[-g^{rr}g_{ux,r}-g_{uu,x}+g^{rx}g_{xx,u}\big]} \,, \\
&& {\textstyle \Gamma^r_{xx} = -\Theta g^{rr}g_{xx}-g_{ux||x}+\frac{1}{2}g_{xx,u}} \,, \\
&& {\textstyle \Gamma^u_{rr}=\Gamma^u_{ru}=\Gamma^u_{rx} = 0} \,, \\
&& {\textstyle \Gamma^u_{uu} = \frac{1}{2}g_{uu,r}} \,, \\
&& {\textstyle \Gamma^u_{ux} = \frac{1}{2}g_{ux,r}} \,, \\
&& {\textstyle \Gamma^u_{xx} = \Theta g_{xx}} \,, \\
&& {\textstyle \Gamma^x_{rr} = 0} \,, \\
&& {\textstyle \Gamma^x_{ru} = \frac{1}{2}g^{xx}g_{ux,r}} \,, \\
&& {\textstyle \Gamma^x_{rx} = \Theta} \,, \\
&& {\textstyle \Gamma^x_{uu} = \frac{1}{2}\big[-g^{rx}g_{uu,r}+g^{xx}(2g_{ux,u}-g_{uu,x})\big]} \,, \\
&& {\textstyle \Gamma^x_{ux} = \frac{1}{2}\big[-g^{rx}g_{ux,r}+g^{xx}g_{xx,u}\big]} \,, \\
&& {\textstyle \Gamma^x_{xx} = -\Theta g^{rx}g_{xx}+\,^{S}\Gamma^x_{xx}} \,, \label{ChristoffelEnd}
\end{eqnarray}
where
\begin{align} \label{sGamma}
\,^{S}\Gamma^x_{xx}\equiv\tfrac{1}{2}g^{xx}g_{xx,x} = -\frac{G_{,x}}{G}
\end{align}
is the Christoffel symbol with respect to the only spatial coordinate $x$, i.e.,  coefficient of the covariant derivative on the transverse 1-dimensional space spanned by~$x$.

The non-vanishing Riemann curvature tensor components are then
\begin{eqnarray}
&& R_{rxrx} = {\textstyle -\big(\Theta_{,r}+\Theta^2\big)g_{xx}} \,, \\
&& R_{rxru} = {\textstyle -\frac{1}{2}g_{ux,rr}+\frac{1}{2}\Theta g_{ux,r}} \,, \\
&& R_{ruru} = {\textstyle -\frac{1}{2}g_{uu,rr}+\frac{1}{4}g^{xx}(g_{ux,r})^2} \,, \\
&& R_{rxux} = {\textstyle \frac{1}{2}g_{ux,r||x}+\frac{1}{4}(g_{ux,r})^2-g_{xx}\Theta_{,u}
-\frac{1}{2}\Theta\big(g_{xx,u}+g_{xx}g_{uu,r}\big)} \,, \\
&& R_{ruux} = {\textstyle g_{u[u,x],r}+\frac{1}{4}g^{rx}(g_{ux,r})^2-\frac{1}{4}g^{xx}g_{xx,u}g_{ux,r}} \nonumber \\
&& \hspace{15.0mm} {\textstyle +\Theta\big(g_{ux,u}-\frac{1}{2}g_{uu,x}-\frac{1}{2}g_{ux}g_{uu,r}\big)} \,, \\
&& R_{uxux} = {\textstyle -\frac{1}{2}(g_{uu})_{||xx}+g_{ux,u||x}-\frac{1}{2}g_{xx,uu}+\frac{1}{4}g^{rr}(g_{ux,r})^2} \nonumber \\
&& \hspace{15.0mm} {\textstyle -\frac{1}{2}g_{uu,r}e_{xx}+\frac{1}{2}g_{uu,x}g_{ux,r}-\frac{1}{2}g^{rx}g_{xx,u}g_{ux,r}+\frac{1}{4}g^{xx}(g_{xx,u})^2} \nonumber \\
&& \hspace{15.0mm} {\textstyle -\frac{1}{2}\Theta g_{xx}\big[g^{rr}g_{uu,r}+g_{uu,u}-g^{rx}(2g_{ux,u}-g_{uu,x})\big]} \,.
\end{eqnarray}
Finally, the components of the Ricci tensor are
\begin{eqnarray}
&& R_{rr} = {\textstyle -\big(\Theta_{,r}+\Theta^2\big)} \,, \label{Ricci rr}\\
&& R_{rx} = {\textstyle -\frac{1}{2}g_{ux,rr}+\frac{1}{2}\Theta g_{ux,r}+\big(\Theta_{,r}+\Theta^2\big)g_{ux}} \,, \label{Ricci rp} \\
&& R_{ru} = {\textstyle -\frac{1}{2}g_{uu,rr}+\frac{1}{2}g^{rx}g_{ux,rr}+\frac{1}{2}g^{xx}\big(g_{ux,r||x}+(g_{ux,r})^2\big)} \nonumber \\
&& \hspace{12.0mm} {\textstyle -\Theta_{,u}-\frac{1}{2}\Theta\big(g^{xx}g_{xx,u}+g^{rx}g_{ux,r}+g_{uu,r}\big)} \,, \label{Ricci ru} \\
&& R_{xx} = {\textstyle -g_{xx}\,g^{rr}\big(\Theta_{,r} +\Theta^2 \big) +2g_{xx}\big(\Theta_{,u}-g^{rx}\Theta_{,x}\big)+2g_{ux}\Theta_{,x}-f_{xx}} \nonumber \\
&& \hspace{12.0mm} {\textstyle +\Theta\big[2g_{ux||x}+2g_{ux,r}g_{ux}+g_{xx}\big(g_{uu,r}-2g^{rx}g_{ux,r}\big)
-2e_{xx}\big]} \,, \label{Ricci pq} \\
&& R_{ux} = {\textstyle -\frac{1}{2}g^{rr}g_{ux,rr}-\frac{1}{2}g_{uu,rx}+\frac{1}{2}g_{ux,ru}
-\frac{1}{2}g^{rx}\big[g_{ux,r||x}+(g_{ux,r})^2\big]} \nonumber \\
&& \hspace{12.0mm} {\textstyle +g^{xx}\big(\frac{1}{2}g_{ux,r}g_{ux||x}-\frac{1}{2}e_{xx}g_{ux,r}\big)+g_{ux}\Theta_{,u}}\nonumber \\
&& \hspace{12.0mm} {\textstyle +\Theta\big[g_{ux}g_{uu,r}-\frac{1}{2}(g_{uu}g_{ux,r}-g_{uu,x})-g_{ux,u}
+\frac{1}{2}g^{rx}g_{ux,r}g_{ux}+\frac{1}{2}g^{rx}g_{xx,u}\big]} \,, \label{Ricci up} \\
&& R_{uu} = {\textstyle -\frac{1}{2}g^{rr}g_{uu,rr}-g^{rx}g_{uu,rx}-\frac{1}{2}g^{xx}e_{xx}g_{uu,r}+g^{rx}g_{ux,ru}-\frac{1}{2}g^{xx}g_{xx,uu}} \nonumber \\
&& \hspace{12.0mm} {\textstyle +g^{xx}(g_{ux,u||x}-\frac{1}{2}g_{uu||xx})+\frac{1}{2}(g^{rr}g^{xx}-g^{rx}g^{rx})(g_{ux,r})^2+\frac{1}{2}g^{xx}g_{ux,r}g_{uu,x} +\frac{1}{4}(g^{xx} g_{xx,u})^2} \nonumber \\
&& \hspace{12.0mm} {\textstyle
+\frac{1}{2}\Theta\big[-g^{rx}(2g_{ux,u}-g_{uu,x}-g_{ux}g_{uu,r})+g_{uu}g_{uu,r}-g_{uu,u}\big]} \,, \label{Ricci uu}
\end{eqnarray}
and the Ricci scalar is
\begin{eqnarray}
&& R = {\textstyle g_{uu,rr}-2g^{rx}g_{ux,rr}-2g^{xx}g_{ux,r||x}-\frac{3}{2}g^{xx}(g_{ux,r})^2} \nonumber \\
&& \hspace{8.0mm} {\textstyle +2\Theta_{,r}\,g_{uu}+4\Theta_{,u} +2\Theta^2g_{uu} +\Theta(2g_{uu,r}+2g^{rx}g_{ux,r}+2g^{xx}g_{xx,u})} \,.
\end{eqnarray}
The symbol ${\,_{||}}$ denotes the covariant derivative with respect to $g_{xx}$\,:
\begin{eqnarray}
g_{ux||x} \rovno g_{ux,x}-g_{ux}\,^{S}\Gamma^{x}_{xx}  \,, \\
g_{ux,r||x} \rovno g_{ux,rx}-g_{ux,r}\,^{S}\Gamma^{x}_{xx}  \,, \\
g_{ux,u||x} \rovno g_{ux,ux}-g_{ux,u}\,^{S}\Gamma^{x}_{xx}  \,, \\
(g_{uu})_{||xx} \rovno g_{uu,xx}-g_{uu,x}\,^{S}\Gamma^{x}_{xx}\,,
\end{eqnarray}
where ${e_{xx}}$ and ${f_{xx}}$ are convenient shorthands defined as
\begin{eqnarray}
e_{xx} \!\!\!\!& \equiv &\!\!\!\! g_{u{x||x}}- {\textstyle \frac{1}{2}}g_{xx,u} \,, \label{exx}\\
f_{xx} \!\!\!\!& \equiv &\!\!\!\! g_{ux,r||x}+ {\textstyle \frac{1}{2}}(g_{ux,r})^2 \label{fxx}\,.
\end{eqnarray}
The expressions (\ref{Ricci rr})--(\ref{Ricci uu}) of the Ricci tensor enable us to write explicitly the gravitational field equations for any  ${D=3}$ Kundt or Robinson--Trautman spacetime.


\newpage









\begin{thebibliography}{10}


\bibitem{PodolskySvarcMaeda:2019}
J.~Podolsk\'y, R.~\v{S}varc and H.~Maeda,
All solutions of Einstein's equations in 2+1 dimensions: $\Lambda$-vacuum, pure radiation, or gyratons,
Class. Quantum Grav. {\bf 36} (2019) 015009 (31pp).
\bibitem{btz1}
M.~Ba\~nados, C.~Teitelboim and J.~Zanelli,
The Black hole in three-dimensional space-time,
Phys. Rev. Lett. {\bf 69} (1992) 1849--51.
\bibitem{btz2}
M.~Ba\~nados, M.~Henneaux, C.~Teitelboim and J.~Zanelli,
Geometry of the (2+1) black hole,
Phys. Rev.~D {\bf 48} (1993) 1506--15.
\bibitem{chargedbtz}
C.~Martinez, C.~Teitelboim and J.~Zanelli,
Charged rotating black hole in three space-time dimensions,
Phys. Rev.~D {\bf 61} (2000) 104013 (8pp).
\bibitem{carlip}
S.~Carlip, {\it Quantum Gravity in 2+1 Dimensions}
(Cambridge University Press, Cambridge, 2003).
\bibitem{GarciaDiaz:2017}
A.~A.~Garc\'\i a-D\'\i az, {\em Exact Solutions in Three-Dimensional Gravity}
(Cambridge University Press, Cambridge, 2017).
\bibitem{KrongosTorre:2017}
D.~S.~Krongos and C.~G.~Torre, Rainich conditions in (2+1)-dimensional gravity,
J.~Math. Phys. {\bf 58}, (2017) 012501.
\bibitem{ChowPopeSezgin:2009}
D.~D.~K.~Chow, C.~N.~Pope and E.~Sezgin,
Kundt spacetimes as solutions of topological massive gravity,
Class. Quantum Grav. {\bf 27} (2010) 105002 (19pp).
\bibitem{Kundt:1961}
W.~Kundt, The plane-fronted gravitational waves,
Z.~Phys. {\bf 163} (1961) 77--86.
\bibitem{Stephani:2003}
H.~Stephani, D.~Kramer, M.~MacCallum, C.~Hoenselaers and E.~Herlt, {\em Exact Solutions of Einstein's Field Equations}
(Cambridge University Press, Cambridge, 2003).
\bibitem{GriffithsPodolsky:2009}
J.~B.~Griffiths and J.~Podolsk\'{y}, {\em Exact Space-Times in Einstein's General Relativity}
(Cambridge University Press, Cambridge, 2009).
\bibitem{PodolskyZofka:2009}
J.~Podolsk\'{y} and M.~\v{Z}ofka,
General Kundt spacetimes in higher dimensions,
Class. Quantum Grav. {\bf 26} (2009) 105008 (18pp).
\bibitem{OrtaggioPravdaPravdova:2013}
M.~Ortaggio, V.~Pravda and A.~Pravdov\'a,
Algebraic classification of higher dimensional spacetimes based on null alignment,
Class. Quantum Grav. {\bf 30} (2013)  013001 (57pp).
\bibitem{RobTra60}
I.~Robinson and A.~Trautman, Spherical gravitational waves,
Phys. Rev. Lett. {\bf 4} (1960) 431--2.
\bibitem{RobTra62}
I.~Robinson and A.~Trautman, Some spherical gravitational waves in general relativity,
Proc. Roy. Soc.~A {\bf 265} (1962) 463--73.
\bibitem{PodOrt06}
J.~Podolsk\'{y} and M.~Ortaggio, Robinson-Trautman spacetimes in higher dimensions,
Class. Quantum Grav. {\bf 23} (2006) 5785--97.
\bibitem{OrtPodZof08}
M.~Ortaggio, J.~Podolsk\'{y} and M.~\v{Z}ofka, Robinson-Trautman spacetimes with an electromagnetic field in higher dimensions,
Class. Quantum Grav. {\bf 25} (2008) 025006 (18pp).
\bibitem{OrtaggioPodolskyZofka:2015}
M.~Ortaggio, J.~Podolsk\'{y} and M.~\v{Z}ofka,
Static and radiating p-form black holes in the higher dimensional Robinson-Trautman class,
JHEP {\bf 1502} (2015) 045 (39pp).
\bibitem{MaedaMartinez:2020}
H.~Maeda and C.~Mart\'\i nez,
Energy conditions in arbitrary dimensions,
Prog. Theor. Exp. Phys. {\bf 2020} (2020) 043E02 (35pp).
\bibitem{PodSva13}
J.~Podolsk\'{y} and R.~\v{S}varc, Explicit algebraic classification of Kundt geometries in any dimension,
Class. Quantum Grav. {\bf 30} (2013) 125007 (25pp).
\bibitem{PodSva15}
J.~Podolsk\'{y} and R.~\v{S}varc, Algebraic structure of Robinson-Trautman and Kundt geometries in arbitrary dimension,
Class. Quantum Grav. {\bf 32} (2015) 015001 (34pp).
\bibitem{PodSva16}
J.~Podolsk\'{y} and R.~\v{S}varc, Algebraic classification of Robinson-Trautman spacetimes,
Phys. Rev.~D {\bf 94} (2016) 064043 (15pp).
\bibitem{PlebanHacyan:1979}
J.~F.~Pleba\'nski and S.~Hacyan, Some exceptional electrovac type D metrics with cosmological constant,
J.~Math. Phys. {\bf 20} (1979) 1004--10.
\bibitem{Peldan:1993}
P.~Peldan, Unification of gravity and Yang-Mills theory in (2+1)-dimensions,
Nucl. Phys.~B {\bf 395} (1993) 239--62.
\bibitem{GottAlpert:1984}
J.~R.~Gott and M.~Alpert, General relativity in a (2+1)-dimensional spacetime,
Gen. Rel. Grav. {\bf 16} (1984) 243--7.
\bibitem{GottSimonAlpert:1986}
J.~R.~Gott, J.~Z.~Simon and M.~Alpert, General relativity in a (2+1)-dimensional spacetime: An electrically charged solution,
Gen. Rel. Grav. {\bf 18} (1986) 1019--35.
\bibitem{DeserMazur:1985}
S.~Deser and P.~O.~Mazur, Static solutions in ${D=3}$ Einstein-Maxwell theory,
Class. Quantum Grav. {\bf 2} (1985) L51--6.
\bibitem{Melvin:1986}
M.~A.~Melvin, Exterior solutions for electric and magnetic stars in 2+1 dimensions,
Class. Quantum Grav. {\bf 3} (1986) 117--31.
\bibitem{Garcia:2009}
A.~A.~Garc\'\i a, Three-dimensional stationary cyclic symmetric Einstein-Maxwell solutions; black holes,
Ann. of Phys. {\bf 324} (2009) 2004--50.
\bibitem{GarciaDiaz:2013}
A.~A.~Garc\'\i a-D\'\i az, Three dimensional stationary cyclic symmetric Einstein-Maxwell solutions; energy, mass, momentum, and algebraic tensors characteristics, arXiv:1307.6652 [gr-qc].
\bibitem{Brinkmann:1925}
H.~W.~Brinkmann, Einstein spaces which are mapped conformally on each other,
Math.~Ann. {\bf 94} (1925) 119--45.
\bibitem{Kruchkovich:1955}
G.~I.~Kruchkovich, Invariant criteria of spaces $V_3$ with the group of motions~$G_4$,
Uspekhi Mat. Nauk, {\bf 10} (1955) 129--136.
\bibitem{Chow:2019}
D.~D.~K.~Chow, Characterization of three-dimensional Lorentzian metrics that admit four Killing vectors, arXiv:1903.10496 [gr-qc]
\bibitem{Clement:1993}
G.~Cl\'ement, Classical solutions in three-dimensional Einstein-Maxwell cosmological gravity,
Class. Quantum Grav. {\bf 10} (1993) L49--L54.

\end{thebibliography}
\end{document}